\documentclass[a4paper,10pt]{article} 

\usepackage{lipsum}
\usepackage{amsfonts}
\usepackage{graphicx}
\usepackage{epstopdf}
\usepackage{algorithmic}
\ifpdf
  \DeclareGraphicsExtensions{.eps,.pdf,.png,.jpg}
\else
  \DeclareGraphicsExtensions{.eps}
\fi

\usepackage[utf8]{inputenc}
\usepackage{amsmath}
\usepackage{latexsym}
\usepackage{amsfonts}
\usepackage{amssymb}
\usepackage{stmaryrd}
\usepackage{caption}
\usepackage{color}
\usepackage{enumerate}
\usepackage{bbm}
\usepackage{bm}
\usepackage{dsfont}

\usepackage{subcaption}

\usepackage{tikz}
\usepackage{pgfplots}
\usepackage{xcolor}
\usepackage{hyperref}

\usepackage{tikz-cd}
\tikzcdset{ampersand replacement=\&}

\pgfplotsset{plot coordinates/math parser=false}

\definecolor{airforceblue}{rgb}{0.26, 0.34, 0.96}
	
\newtheorem{algoritmo}{\bf Algorithm}

\hypersetup{hidelinks}

\newcommand{\mN}{{\mathcal N}}

\newcommand{\br}{\boldsymbol{r}}

\newcommand{\bv}{\boldsymbol{v}}
\newcommand{\bw}{\boldsymbol{w}}
\newcommand{\bx}{\boldsymbol{x}}
\newcommand{\by}{\boldsymbol{y}}
\newcommand{\bz}{\boldsymbol{z}}
\newcommand{\bC}{\boldsymbol{C}}

\newcommand{\bH}{\boldsymbol{H}} 
\newcommand{\bI}{\boldsymbol{I}} 
\newcommand{\bJ}{\boldsymbol{J}}
\newcommand{\bK}{\boldsymbol{K}} 
 
\newcommand{\bP}{\boldsymbol{P}}
\newcommand{\bQ}{\boldsymbol{Q}}
\newcommand{\bR}{\boldsymbol{R}}
\newcommand{\bS}{\boldsymbol{S}}
\newcommand{\bT}{\boldsymbol{T}}
\newcommand{\bU}{\boldsymbol{U}}
\newcommand{\bV}{\boldsymbol{V}}

\newcommand{\bX}{\boldsymbol{X}}
\newcommand{\bY}{\boldsymbol{Y}}

\newcommand{\btheta}{\boldsymbol{\theta}}

\def\wfigtwo{5.3cm}
\def\hfigtwo{3.4cm}

\newcommand{\Reals}{\mathbb R}      
\newcommand{\Integers}{\mathbb Z}   
\newcommand{\Naturals}{\mathbb N}   
\usepackage[acronym]{glossaries}

\newacronym{AWGN}{AWGN}{addtive white {G}aussian noise}
\newacronym{apf}{APF}{auxiliary particle filter}
\newacronym{ckf}{CKF}{cubature Kalman filter}
\newacronym{ekf}{EKF}{extended Kalman filter}
\newacronym{enkf}{EnKF}{ensemble Kalman filter}
\newacronym{hmm}{HMM}{heterogeneous multi-scale model}
\newacronym{iid}{i.i.d.}{independent and identically distributed}
\newacronym{ipf}{IPF}{island particle filter}
\newacronym{kf}{KF}{Kalman filter}
\newacronym{mse}{MSE}{mean square error}
\newacronym{nhf}{NHF}{nested hybrid filter}
\newacronym{ngf}{NGF}{nested Gaussian filter}
\newacronym{npf}{NPF}{nested particle filter}
\newacronym{nsmc}{NSMC}{nested sequential Monte Carlo}
\newacronym{nmse}{NMSE}{normalized mean square error}
\newacronym{pdf}{pdf}{probability density function}
\newacronym{pmcmc}{PMCMC}{particle Markov chain Monte Carlo}
\newacronym{pf}{PF}{particle filter}
\newacronym{pl}{PL}{particle learning}
\newacronym{pde}{PDE}{partial differential equation}
\newacronym{qmc}{QMC}{quasi Monte Carlo}
\newacronym{rv}{r.v.}{random variable}
\newacronym{rml}{RML}{recursive maximum likelihood}
\newacronym{sir}{SIR}{sequential importance resampling}
\newacronym{smc}{SMC}{sequential Monte Carlo}
\newacronym{smc2}{SMC$^2$}{sequential Monte Carlo square}
\newacronym{sqmc}{SQMC}{sequential quasi-Monte Carlo}
\newacronym{sde}{SDE}{stochastic differential equation}
\newacronym{stpf}{ST-PF}{space-time particle filter}
\newacronym{ukf}{UKF}{unscented Kalman filter}
\newacronym{ut}{UT}{unscented transform}
\newacronym{wrt}{w.r.t.}{with respect to}

\newcommand\blfootnote[1]{%
  \begingroup
  \renewcommand\thefootnote{}\footnote{#1}%
  \addtocounter{footnote}{-1}%
  \endgroup
}

\usepackage{authblk}

\title{Nested smoothing algorithms for inference and tracking of heterogeneous multi-scale state-space systems}

\author[1]{Sara P\'erez-Vieites}
\author[2]{Harold Molina-Bulla}
\author[3]{Joaqu\'{\i}n M\'iguez}

\affil[1]{Department of Electrical Engineering and Automation, Aalto University,
Finland.}
\affil[2,3]{Department of Signal Theory and Communications, Universidad Carlos III de Madrid, Spain.}

\usepackage{amsopn}

\providecommand{\keywords}[1]
{
  \small	
  \textbf{\textit{Keywords---}} #1
}

\begin{document}

\maketitle
\blfootnote{Emails: $^1$sara.perezvieites@aalto.fi, $^2$h.molina@tsc.uc3m.es ,$^3$joaquin.miguez@uc3m.es }






\abstract{
Multi-scale problems, where variables of interest evolve in different time-scales and live in different state-spaces, can be found in many fields of science. Here, we introduce a new recursive methodology for Bayesian inference that aims at estimating the static parameters and tracking the dynamic variables of this kind of systems. Although the proposed approach works in rather general setups, for clarity we analyze the case of a heterogeneous multi-scale model with 3 time-scales (static parameters, slow dynamic state variables and fast dynamic state variables). The proposed scheme, based on a nested filtering methodology of \cite{Perez-Vieites18}, combines three intertwined layers of filtering techniques that approximate recursively the joint posterior probability distribution of the parameters and both sets of dynamic state variables given a sequence of partial and noisy observations. We explore the use of both sequential Monte Carlo schemes and several Kalman filtering techniques in the different layers of the methodology to obtain approximations of the posterior probability distributions of interest. Some numerical results are presented for a stochastic two-scale Lorenz 96 model with unknown parameters.
}



\section{Introduction}
 
\subsection{Motivation}
 
 {	
	Multi-scale problems can be found naturally in many fields of science, such as biology, chemistry, fluid dynamics and material science, where processes at different time and/or spatial scales may be described by diverse laws \cite{Pavliotis08,Weinan11,Weinan03}. Thanks to the improvements in computational power and the need for ever more faithful models of real-world complex systems, the interest in multi-scale modelling techniques has increased in recent years. 
}

 {
	Multi-scale models are aimed at providing an efficient and more accurate representation of a complex system by coupling (sub)models that address different features/attributes at different levels of detail. However, the modeling of such systems is far from straightforward, not only because they are made up of dynamical systems of different characteristics, but also because of often intricate cross dependencies among the relevant physical processes at work \cite{Weinan11}. As a consequence, the problem of tracking the evolution of a multi-scale dynamical system involves the prediction and estimation of several sets of variables that live in different state-spaces and evolve in different time scales. Moreover, the tracking of the variables of interest usually has to be performed from the observation of partial and noisy observations. Efficient recursive algorithms for this task are needed.
}

\subsection{Background}

 {
	The simplest case of a multi-scale problem consists of a system with unknown static parameters and dynamic state variables, since the parameters may be considered as state variables that evolve at a greater time scale. Hence, it is a multi-scale problem with only two time scales. 
	In general, carrying out both parameter estimation and state tracking at once implies several practical and theoretical difficulties. Within this framework, a few well-principled methods have been proposed in the last few years. Two examples of schemes that yield theoretically-guaranteed solutions to this problem are \gls{smc2} \cite{Chopin12} and \gls{pmcmc} \cite{Andrieu10}. They aim at computing the joint posterior probability distribution of all the unknown variables and parameters of the system, which provides all the information to obtain point estimates and also to quantify the estimation error. Unfortunately, both \gls{smc2} and \gls{pmcmc} are batch techniques. In other words, every time a new observation arrives, the whole sequence of observations may have to to be re-processed from scratch in order to update the estimates, leading to a quadratic increase of the computational effort over time. As an alternative, \glspl{npf} \cite{Crisan18bernoulli} apply the same principles as \gls{smc2} in a recursive way. An \gls{npf} estimates both parameters and states using a scheme of two intertwined layers of Monte Carlo methods, one inside the other.
    Here, the first layer approximates the distribution of the parameters given the observations, while the second layer approximates the distribution of the dynamical variables conditional on the parameters and observed data.
 This methodology is better suited for long sequences of observations, however, the use of Monte Carlo in both layers of filters still makes its computational cost prohibitive in high-dimensional problems. {The methods in \cite{Fang23,Perez-Vieites21,Perez-Vieites18} are extensions or modifications of the \gls{npf} that aim at reducing its computational complexity. The order of the layers is reversed in \cite{Fang23}, reformulating the problem such that the parameters are integrated out following the idea of Rao-Blackwellization. 
 }
 \Glspl{nhf}  \cite{Perez-Vieites18} and \glspl{ngf} \cite{Perez-Vieites21} introduce the use of Kalman filtering techniques (i.e., Gaussian approximations of the distributions) in one or both layers of the algorithm, respectively. The introduction of this approximation reduces the computational cost considerably and makes the methodology more appealing for online processing.
}

 {
	In the last few years, other algorithms with nested, or layered, structures (in the vein of SMC$^2$, NPF or NHF) have been proposed in order to address inference in high-dimensional, complex models. The most recent examples are the \gls{stpf} \cite{Beskos17} and the \gls{nsmc} method \cite{Naesseth19}. Both techniques are intended to outperform classical sequential Monte Carlo (SMC) in high dimensional systems. They rely on spatial structures within the state space (a Markov random field in \cite{Naesseth19} and an auto-regressive structure in \cite{Beskos17}) and, therefore, they may be useful to tackle multiple spatial scales. 
}


 {			
	One of the most typical examples of multi-scale systems is the two-scale Lorenz 96 model \cite{Arnold13,Carlu18,Lorenz96}. This is a simplified weather model which includes different spatial and temporal scales. Specifically, it is a heterogeneous multi-scale model, i.e. a model where the macro-scale level description needs to be completed by the data extracted from a micro-scale level \cite{Vanden07}. The coupling of both scales can be handled in two different ways:
	\begin{itemize}
	    \item applying parameterization \cite{Arnold13,Pavliotis08,Vissio18} or 
	    \item using a macro-micro solver \cite{Weinan03,Weinan05}. 
	\end{itemize}
	On one hand, parameterization is used to replace the contribution of the micro-scale level of the model by a simplified process that depends only on the slow state variables of the macro scale. Once the model is simplified to a single scale, many other algorithms can be used to estimate the evolution of the macro-scale process, that is usually the scale of interest.} On the other hand, micro-macro solvers keep both scales and avoid any simplification of the model. Inference in the two-scale Lorenz 96 model has been addressed using algorithms such as particles filters \cite{Lingala12,Miguez15eus,Yeong20}, different Kalman filters \cite{Grooms15,Pulido18,Shen18,Tsuyuki12} or a combination of different methods \cite{Grooms21,Santitissadeekorn15}. 
	 
	
	

\subsection{Contributions}

 {
	In this paper, we propose a generalization of the \gls{nhf} methodology aimed at performing recursive Bayesian inference for a class of heterogeneous multi-scale state-space models \cite{Abdulle2012} that can be numerically approximated with a micro-macro solver.We analyse the case of a Lorenz 96 system that displays three time scales (static parameters, slow dynamic state variables at the macro-scale, and fast dynamic state variables at the micro-scale), but the methodology works in the same way for more general examples (namely, systems with $n$ scales either in time or space). 
}

 
	The new scheme can be described as a three-layer nested smoother that approximates, in a recursive manner, the posterior probability distributions of the parameters and the two sets of state variables given the sequence of available observations. Specifically, we approximate the posterior probability distribution of the parameters in a first layer of computation, the posterior probability distribution of the slow state variables in a second layer, and the posterior probability distribution of the fast state variables in a third layer. The computations on the second layer are conditional on the candidate parameter values generated on the first layer, while the calculations on the third layer are conditional on the candidates drawn at the first and second layers. The inference techniques used in each layer can vary, leading to different computational costs and degrees of accuracy. As examples that illustrate the methodology, we propose two methods. The first one uses SMC algorithms in the first and second layers, intertwined with an unscented Kalman filter (UKF) \cite{Julier00} in the third layer. Similarly, the second method uses a \gls{smc} algorithm in the first layer, but incorporates the use of \glspl{enkf} \cite{Evensen03} and \glspl{ekf} in the second and third layers of the scheme, respectively.

\subsection{Organization}

 {
	The rest of the paper is organized as follows. We introduce the notation used through the paper and state the problem to be addressed in Section \ref{sec:Multi ProblemStatement}. In Section \ref{sec:Multi Optimalfilter}, we describe the optimal smoother for multi-scale systems with static parameters and two sets of dynamic state variables. Two specific methods derived from the general methodology are shown in Section \ref{sec:Multi MultiscaleNestedFilter}. Numerical results for the stochastic two-scale Lorenz 96 model are shown in Section \ref{sec:Multi Example} and conclusions are drawn in Section \ref{sec:Multi Conclusions}.
}

%
\section{Problem Statement} \label{sec:Multi ProblemStatement}

%
\subsection{Notation}

 {
We use lower-case, regular-face letters, e.g., $x$, to denote scalar quantities and bold-face, e.g., $\bx$, to denote vectors. Matrices are represented by bold-face upper-case letters, e.g., $\bX$. We make no notational difference between deterministic quantities and random variables (r.v.'s).
}

 {
If $\bx$ is a $d$-dimensional random vector on $\Reals^d$, we denote its pdf with respect to the Lebesgue measure as $p(\bx)$. This is an argument-wise notation, i.e., if $\bx$ and $\by$ are random vectors, we denote their pdf's as $p(\bx)$ and $p(\by)$ even if they are possibly different functions (when $\bx$ and $\by$ obey different probability laws). Similary, $p(\bx,\by)$ denotes the joint pdf of $\bx$ and $\by$, while $p(\bx|\by)$ is the conditional pdf of $\bx$ given $\by$. This notation, which has been broadly used in the field of particle filtering \cite{Djuric03,Doucet00,Doucet09,Liu98}, is simple yet sufficient to describe the methodologies in this paper.
}

 {
We also resort to a more specific notation for Gaussian pdf's. If $\bx$ is a $d$-dimensional Gaussian random vector with mean $\widehat \bx$ and covariance matrix $\bC>0$ then we can explicitly write the pdf $p(\bx)$ as 
$$
\mN(\bx | \widehat \bx, \bC) := 
\frac{1}{\left( 2\pi \right)^{\frac{d}{2}} |\bC|^{\frac{1}{2}} } 
\exp\left\{
	-\frac{1}{2}\left( 
		\bx - \widehat \bx
	\right)^\top \bC^{-1} \left(
		\bx - \widehat \bx
	\right)
\right\},
$$
where the superscript $^\top$ indicates transposition and $|\bC|$ denotes the determinant of matrix $\bC$.
}

%
\subsection{State space models} \label{subsec:Multi Statespacemodels}

In this chapter we place our attention on state space models that result from the analysis of physical systems that display (intertwined) dynamical features at different time scales. To be specific, let us consider the class of multidimensional additive \glspl{sde} that can be written as
\begin{eqnarray}
 d\bx = f_{\bx}(\bx,\btheta)d\tau + g_{\bx}(\bz,\btheta)d\tau+ \bQ_x d\bv, \label{eqsdex}\\
 d\bz = f_{\bz}(\bx,\btheta)d\tau + g_{\bz}(\bz,\btheta)d\tau + \bQ_z d\bw, \label{eqsdez}
\end{eqnarray}
where 
\begin{itemize}
    \item $\tau$ denotes continuous time, 
    \item $\bx(\tau) \in \Reals^{d_x}$ and $\bz(\tau) \in \Reals^{d_z}$ are the slow and fast states of the system, respectively, 
    \item $f_{\bx} \colon \Reals^{d_x} \times \Reals^{d_\theta} \to \Reals^{d_x}$, $g_{\bx} \colon \Reals^{d_z} \times \Reals^{d_\theta} \to \Reals^{d_x}$, $f_{\bz} \colon \Reals^{d_x} \times \Reals^{d_\theta} \to \Reals^{d_z}$ and $g_{\bz} \colon \Reals^{d_z} \times \Reals^{d_\theta} \to \Reals^{d_z}$ are (possibly nonlinear) transition functions parameterized by a fixed vector of unknown parameters, $\btheta \in \Reals^{d_\theta}$, 
    \item $\bQ_x$ and $\bQ_z$ are known diffusion matrices that control the intensity and covariance of the stochastic perturbations, 
    \item and $\bv(\tau)$ and $\bw(\tau)$ are vectors of independent standard Wiener processes with dimension $d_x$ and $d_z$, respectively.
\end{itemize}

Equations \eqref{eqsdex}--\eqref{eqsdez} do not have closed form solutions for general nonlinear functions $f_{\bx}$, $f_{\bz}$, $g_{\bx}$ and $g_{\bz}$, and they have to be discretized for their numerical integration. In order to handle the slow and fast time scales, we apply a macro-micro solver \cite{Vanden03,Weinan05} that runs an Euler-Maruyama scheme for each set of state variables, albeit with different integration steps. To be specific, we use $\Delta_z$ as the integration step of $\bz$ while $\Delta_x \gg \Delta_z$ is the integration step of $\bx$. Then, we can simulate $\bx$ and $\bz$ using the pair of difference equations
\begin{eqnarray}
    \bx_{t} &=& \bx_{t-1} + \Delta_x(f_{\bx}(\bx_{t-1}, \btheta) +g_{\bx}(\overline{\bz}_{t},\btheta)) + \sqrt{\Delta_x} \bQ_x \bv_{t}, \label{eqIntx}\\
    \bz_n &=& \bz_{n-1} + \Delta_z(f_{\bz}(\bx_{\lfloor\frac{n-1}{h}\rfloor}, \btheta) +g_{\bz}({\bz}_{n-1},\btheta)) + \sqrt{\Delta_z} \bQ_z \bw_n, \label{eqIntz}
\end{eqnarray}
where $\bx_{t} \approx \bx(t\Delta_x)$ and $\bz_n \approx \bz(n\Delta_z)$ are the state signals, $t \in \Naturals$ denotes discrete time in the time scale of the slow variables, $n \in \Naturals$ denotes discrete time in the fast time scale, $h = \frac{\Delta_x}{\Delta_z} \in \Integers^+$ is the number of fast steps (in the scale of $\bz$) per slow step (in the scale of $\bx$), $\bv_{t}$ and $\bw_n$ are Gaussian \glspl{rv} of zero mean and covariance matrices $\boldsymbol{I}_{d_x}$ and $\boldsymbol{I}_{d_z}$ respectively, and $\overline{\bz}_{t}$ is an average of the fast signal computed as
\begin{equation}
	\overline{\bz}_{t} = \frac{1}{h} \sum_{i = h(t-1)+1}^{ht} \bz_i. \label{eqbarz}
\end{equation}

We assume that the available observations may be directly related to both sets of state variables $\bx_{t}$ and $\bz_n$, but only in the (slow) time scale of $\bx$. To be specific, the $t$-th observation is a $d_y$-dimensional \gls{rv}, $\by_{t} \in \Reals^{d_y}$, which we model as
\begin{equation}
    \by_{t} = l(\bz_{ht}, \bx_{t}, \btheta) + \br_{t}, \label{eqObsY}
\end{equation}
where $l \colon \Reals^{d_z} \times \Reals^{d_x} \times \Reals^{d_\theta} \to \Reals^{d_y}$ is a transformation that maps the states into the observation space, and $\br_{t}$ is a zero-mean observational-noise vector with covariance matrix $\bR$.

%
\subsection{Inference} \label{subsec:Multi Modelinference}

We aim at performing \textit{joint} Bayesian estimation of the parameters, $\btheta$, and all states, $\bx$ and $\bz$, for the state-space model described by Eqs. \eqref{eqIntx}--\eqref{eqIntz} and \eqref{eqObsY}. Typically, the three vectors of unknowns are tightly coupled. The estimation of the fixed parameters is necessary to track both sets of state variables and, at the same time, tracking the slow state variables is needed for predicting the time evolution of the fast states and vice versa.

While in many practical applications one is typically interested in filtering, i.e., the computation of the posterior \gls{pdf} of $\btheta$, $\bx_{t}$ and $\bz_n$ (for $n=ht$) given the data sequence $\by_{1:t} = \{ \by_1, \by_2, \ldots, \by_{t} \}$, we find more convenient to tackle the smoothing \gls{pdf} $p(\bz_{h(t-1)+1:ht}, \bx_{0:t}, \btheta | \by_{1:t})$. Using the chain rule, we can factorize the latter density as
\begin{eqnarray}
p(\bz_{h(t-1)+1:ht}, \bx_{0:t}, \btheta | \by_{1:t}) &=& 
p(\bz_{h(t-1)+1:ht} | \bx_{0:t}, \by_{1:t}, \btheta)  p(\bx_{0:t} | \by_{1:t}, \btheta)  p(\btheta | \by_{1:t}), \quad \quad  \label{eqjpdf}
\end{eqnarray}
where we identify the three key conditional distributions that we seek to compute (or approximate). Each one of these \glspl{pdf} can be handled in a different \textit{layer} of computation. Hence, we aim at designing a nested inference algorithm (in the vein of \cite{Perez-Vieites18}) with three layers. In the first layer we compute $p(\btheta | \by_{1:t})$, in the second one we obtain $p(\bx_{0:t} | \by_{1:t}, \btheta)$, and in the third layer we tackle $p(\bz_{h(t-1)+1:ht} | \bx_{0:t}, \by_{1:t}, \btheta)$.

Hereafter we describe the methodology for the optimal (yet impractical) calculation of the posterior \gls{pdf} in Eq. \eqref{eqjpdf} as well as two approximate numerical solutions that admit feasible computational implementations.



%
\section{Optimal nested smoother} \label{sec:Multi Optimalfilter}

We introduce the optimal nested smoothing algorithm, consisting of three layers, that computes each of the \glspl{pdf} in Eq. \eqref{eqjpdf}. As a result, we obtain the posterior smoothing density $p(\bz_{h(t-1)+1:ht}, \bx_{0:t}, \btheta | \by_{1:t})$ which, in turn, can be used to compute the optimal filtering \gls{pdf}, $p(\bz_{ht}, \bx_{t}, \btheta | \by_{1:t})$, by marginalization if necessary. When the exact computations demanded by this algorithm are not feasible (for general nonlinear and/or non-Gaussian dynamical systems) it serves as a template for approximate numerical schemes, as shown in Section \ref{sec:Multi MultiscaleNestedFilter}.

%
\subsection{First layer: static parameters}

The aim of this layer is to compute the posterior \gls{pdf} of the parameters, $p(\btheta | \by_{1:t})$, recursively. We assume that the \textit{a priori} density $p(\btheta)$ is known.

At time $t$, assume that $p(\btheta  | \by_{1:t-1})$ has been calculated. When a new observation, $\by_{t}$, is obtained, we need to compute the likelihood $p(\by_{t} | \by_{1:t-1}, \btheta)$ in order to obtain the posterior \gls{pdf} of $\btheta$ at time $t$ as
\begin{equation}
p(\btheta | \by_{1:t}) \propto p(\by_{t} | \by_{1:t-1}, \btheta) p(\btheta | \by_{1:t-1}). \label{eq:Multi postparam 1st}
\end{equation}
However, the parameter likelihood $p(\by_{t} | \by_{1:t-1}, \btheta)$ cannot be computed directly. Instead, we decompose it as
\begin{equation}
p(\by_{t} | \by_{1:t-1}, \btheta) = \int p(\by_{t} | \bx_{0:t}, \by_{1:t-1}, \btheta) p(\bx_{0:t} | \by_{1:t-1}, \btheta) d\bx_{0:t}, \label{eq:Multi likelihood 1st}
\end{equation}
where $p(\by_{t} | \bx_{0:t}, \by_{1:t-1}, \btheta)$ and $p(\bx_{0:t} | \by_{1:t-1}, \btheta)$ are the likelihood and the predictive \gls{pdf} of the state sequence $\bx_{0:t}$, respectively, conditional on the previous observations and the parameters. These \glspl{pdf} are computed in the second layer of the algorithm.

\subsection{Second layer: slow states}

Computations in this layer are conditional on the parameter vector $\btheta$. We seek to compute the predictive density $p(\bx_{0:t} |\by_{1:t-1}, \btheta)$, the likelihood $p(\by_{t} | \bx_{0:t}, \by_{1:t-1}, \btheta)$, and the smoothing posterior $p(\bx_{0:t} |\by_{1:t}, \btheta)$.  The first two are needed in the first layer --see Eq. \eqref{eq:Multi likelihood 1st}. We assume that the prior density $p(\bx_0)$ is known and the posterior \gls{pdf} of the slow states at time $t-1$ (conditional on $\btheta$), $p(\bx_{0:t-1}  | \by_{1:t-1}, \btheta)$, is available at time $t$.

We first seek the predictive density of $\bx_{0:t}$, namely,
\begin{equation}
p(\bx_{0:t}| \by_{1:t-1},\btheta) = p(\bx_{t} | \bx_{0:t-1}, \by_{1:t-1},\btheta) p(\bx_{0:t-1}| \by_{1:t-1},\btheta), \label{eq:Multi predx 2nd}
\end{equation}
which is obtained recursively from the posterior at time $t-1$, $p( \bx_{0:t-1}| \by_{1:t-1}, \btheta)$, but requires the evaluation of the marginal density $p(\bx_{t} | \bx_{0:t-1}, \by_{1:t-1},\btheta)$. The latter is not directly available. It has to be computed as an integral \gls{wrt} the fast state variables. In particular 
\begin{equation}
p(\bx_{t} | \bx_{0:t-1}, \by_{1:t-1}, \btheta) = \int p(\bx_{t} | \bz_{h(t-1)+1:ht}, \bx_{0:t-1}, \by_{1:t-1}, \btheta) 
 p(\bz_{h(t-1)+1:ht} |\bx_{0:t-1}, \by_{1:t-1}, \btheta) d\bz_{h(t-1)+1:ht}. \label{eq:Multi transitionx 2nd} 
\end{equation}
The two densities in the integrand of Eq. \eqref{eq:Multi transitionx 2nd}, which involve the fast state variables $\bz_{h(t-1)}, \ldots, \bz_{ht}$, are calculated in the third layer. Recall that $h = \frac{\Delta_x}{\Delta_z}$ is the number of discrete-time steps of the fast states per each single time step of the slow variables (i.e., the $\bz_n$'s are $h$ time faster than the $\bx_{t}$'s).

As for the likelihood, when $\by_{t}$ becomes available we update the posterior density of $\bx_{0:t}$ (conditional on $\btheta$) as
\begin{equation}
p(\bx_{0:t} |\by_{1:t}, \btheta) \propto p(\by_{t} | \bx_{0:t}, \by_{1:t-1}, \btheta) p(\bx_{0:t} | \by_{1:t-1}, \btheta). \label{eq:Multi postx 2nd}
\end{equation}
In the equation above, the likelihood $p(\by_{t} | \bx_{0:t}, \by_{1:t-1}, \btheta)$ can be computed as an integral \gls{wrt} the fast state variables, specifically,
\begin{eqnarray}
p(\by_{t} | \bx_{0:t}, \by_{1:t-1}, \btheta) &=& \int p(\by_{t} | \bz_{h(t-1)+1:ht}, \bx_{0:t}, \by_{1:t-1}, \btheta) 
 p(\bz_{h(t-1)+1:ht} |\bx_{0:t}, \by_{1:t-1}, \btheta) d\bz_{h(t-1)+1:ht} \nonumber \\
&=& \int p(\by_{t} | \bz_{ht}, \bx_{t}, \btheta) p(\bz_{h(t-1)+1:ht} |\bx_{0:t}, \by_{1:t-1}, \btheta) d\bz_{h(t-1)+1:ht}.   \label{eq:Multi likelihood 2nd}
\end{eqnarray}
The likelihood function $p(\by_{t} | \bz_{ht}, \bx_{t}, \btheta)$ can be obtained directly from the state-space model described by Eqs. \eqref{eqIntx}--\eqref{eqObsY}, while the conditional \gls{pdf} of the subsequence $\bz_{h(t-1)+1:ht}$ can be further decomposed as 
\begin{equation}
p(\bz_{h(t-1)+1:ht} |\bx_{0:t}, \by_{1:t-1}, \btheta) =\frac{ p(\bx_{t} | \bz_{h(t-1)+1:ht}, \bx_{t-1}, \btheta)}{p( \bx_{t}| \bx_{0:t-1}, \by_{1:t-1}, \btheta)} 
\times 
p( \bz_{h(t-1)+1:ht}| \bx_{0:t-1}, \by_{1:t-1}, \btheta). \label{eq:Multi predz weird 2nd}
\end{equation} 
Both the likelihood $p(\by_{t} | \bx_{0:t}, \by_{1:t-1}, \btheta)$ of Eq. \eqref{eq:Multi likelihood 2nd} and the {predictive density $p( \bz_{h(t-1)+1:ht}| \bx_{0:t-1}, \by_{1:t-1}, \btheta)$ of Eq. \eqref{eq:Multi predz weird 2nd}} are explicitly computed in the third layer.

%
\subsection{Third layer: fast states}

Computations on this layer are conditional on the parameter vector $\btheta$ and the sequence of slow states $\bx_{0:t}$. In particular, we seek to compute the conditional posterior \glspl{pdf} of $\bz_{h(t-1)+1:ht}$, including the predictive densities,
$$
p(\bz_{h(t-1)+1:ht} | \bx_{0:t-1}, \by_{1:t-1}, \btheta) \quad
\text{and} \quad
p(\bz_{h(t-1)+1:ht} | \bx_{0:t}, \by_{1:t-1}, \btheta),
$$ 
as well as the updated density $p(\bz_{h(t-1)+1:ht} | \bx_{0:t}, \by_{1:t}, \btheta)$. We also evaluate the plain likelihood function $p(\by_{t} |  \bz_{ht}, \bx_{t}, \btheta)$. We assume that the prior \gls{pdf} of the fast states, $p(\bz)$, is known and the posterior up to time $t-1$, $p(\bz_{h(t-2)+1:h(t-1)} | \bx_{t-1}, \by_{1:t-1}, \btheta)$, is available to enable recursive computations.

The first predictive \gls{pdf} is computed recursively from the posterior up to time $t-1$ as the integral
\begin{eqnarray}
p(\bz_{h(t-1)+1:ht} | \bx_{0:t-1}, \by_{1:t-1},\btheta) &=& \int p(\bz_{h(t-1)+1:ht} | \bz_{h(t-2)+1:h(t-1)},\bx_{0:t-1}, \by_{1:t-1},\btheta) \times \nonumber
\\
&\quad&\times p(\bz_{h(t-2)+1:h(t-1)} | \bx_{0:t-1}, \by_{1:t-1}, \btheta) d\bz_{h(t-2)+1:h(t-1)} \nonumber 
\\
&=& \int p(\bz_{h(t-1)+1:ht} | \bz_{h(t-1)},\bx_{t-1}, \btheta) \times \nonumber 
\\
&\quad&\times p(\bz_{h(t-2)+1:h(t-1)} | \bx_{0:t-1}, \by_{1:t-1}, \btheta) d\bz_{h(t-2)+1:h(t-1)}, \label{eq:Multi predz 3rd}
\end{eqnarray}
where the transition \gls{pdf} $p(\bz_{h(t-1)+1:ht} | \bz_{h(t-1)},\bx_{t-1}, \btheta)$ is obtained immediately by iterating Eq. \eqref{eqIntx} in the state-space model $h$ times {and $p(\bz_{h(t-2)+1:h(t-1)} | \bx_{0:t-1}, \by_{1:t-1}, \btheta) d\bz_{h(t-2)+1:h(t-1)}$ is the posterior \gls{pdf} of the fast states in the previous time step. Besides,} the second predictive density, $p(\bz_{h(t-1)+1:ht} | \bx_{0:t}, \by_{1:t-1}, \btheta)$, is obtained by substituting the first predictive \gls{pdf} of Eq. \eqref{eq:Multi predz 3rd} into Eq. \eqref{eq:Multi predz weird 2nd}\footnote{Note that, in Eq. \eqref{eq:Multi predz weird 2nd}, the density $p( \bx_{t}| \bx_{0:t-1}, \by_{1:t-1}, \btheta)$ is the normalization constant for the conditional \gls{pdf} $p( \bz_{h(t-1)+1:ht}| \bx_{0:t-1}, \by_{1:t-1}, \btheta)$, while $p(\bx_{t} | \bz_{h(t-1)+1:ht}, \bx_{t-1}, \btheta)$ results from the iteration of Eq. \eqref{eqIntx}.}.

Finally, when the observation $\by_{t}$ becomes available, we compute the plain likelihood $p(\by_{t} |  \bz_{ht}, \bx_{t}, \btheta)$ (from Eq. \eqref{eqObsY} in the state-space model) and then update the conditional posterior \gls{pdf} of the fast state variables, namely,
\begin{equation}
p(\bz_{h(t-1)+1:ht} | \bx_{0:t}, \by_{1:t}, \btheta) = \frac{ p(\by_{t} | \bz_{ht},\bx_{t}, \btheta) p(\bx_{t} |\bz_{h(t-1)+1:ht}, \bx_{t-1}, \btheta)}{p(\by_{t} | \bx_{0:t}, \by_{1:t-1}, \btheta) p(\bx_{t} |\bx_{0:t-1},\by_{1:t-1}, \btheta)} \times 
 p(\bz_{h(t-1)+1:ht} | \bx_{0:t-1}, \by_{1:t-1}, \btheta) \nonumber
\end{equation}
or, simply,
\begin{equation}
p(\bz_{h(t-1)+1:ht} | \bx_{0:t}, \by_{1:t}, \btheta) \propto ~p(\by_{t} | \bz_{ht},\bx_{t}, \btheta) p(\bx_{t} |\bz_{h(t-1)+1:ht}, \bx_{t-1}, \btheta) 
 p(\bz_{h(t-1)+1:ht} | \bx_{0:t-1}, \by_{1:t-1}, \btheta) \label{eq:Multi postz 3rd}
\end{equation}
if we skip the normalization constant that is typically not needed explicitly for numerical implementations.

%
\subsection{Outline of the optimal nested smoother}
The optimal nested smoother uses each layer of computation to track a subset of \glspl{rv} that evolve over their own time scale, by computing the corresponding predictive and updated \glspl{pdf} (when observations are collected), as well as the necessary likelihoods. To be specific:
\begin{itemize}
	
	\item The third layer tracks the fast state variables, $\bz_n$, and computes the predictive \gls{pdf}  $p(\bz_{h(t-1)+1:ht} | \bx_{0:t-1}, \by_{1:t-1}, \btheta)$ {of Eq. \eqref{eq:Multi predz 3rd}} and the likelihood $p(\by_{t} |  \bz_{ht}, \bx_{t}, \btheta)$. They are used to track the conditional posterior density $p(\bz_{h(t-1)+1:ht} | \bx_{0:t}, \by_{1:t}, \btheta)$ {of Eq. \eqref{eq:Multi postz 3rd}.	}
	
	\item The second layer takes the \gls{pdf} $p(\bz_{h(t-1)+1:ht} | \bx_{0:t-1}, \by_{1:t-1}, \btheta)$ and the likelihood $p(\by_{t} |  \bz_{ht}, \bx_{t}, \btheta)$ in order to compute the predictive \gls{pdf} $p(\bx_{0:t}| \by_{1:t-1},\btheta)$ in Eq. \eqref{eq:Multi predx 2nd} and the likelihood $p(\by_{t} | \bx_{0:t}, \by_{1:t-1}, \btheta)$ in Eq. \eqref{eq:Multi likelihood 2nd}. These are used to track the posterior \gls{pdf} of the slow state, $p(\bx_{0:t} |\by_{1:t}, \btheta)$, {of Eq. \eqref{eq:Multi postx 2nd}}.	
	
	\item The first layer takes the \glspl{pdf} $p(\bx_{0:t}| \by_{1:t-1},\btheta)$ and $p(\by_{t} | \bx_{0:t}, \by_{1:t-1}, \btheta)$ to track the posterior \gls{pdf} of the parameters, $p(\btheta|\by_{1:t})$, {of Eq. \eqref{eq:Multi postparam 1st}}.	
	
\end{itemize}
Finally, the three conditional posterior \glspl{pdf} are needed to compute the joint smoothing density $p(\bz_{h(t-1)+1:ht}, \bx_{0:t}, \btheta | \by_{1:t})$ in Eq. \eqref{eqjpdf}.

Figure \ref{figscheme} is a schematic representation of the optimal smoother, which displays each layer in a different column. Most of the \glspl{pdf} that need to be computed are included in this scheme, showing the dependencies among them with arrows. 
For example, the predictive \gls{pdf} $p(\bz_{h(t-1)+1:ht} | \bx_{0:t-1}, \by_{1:t-1}, \btheta)$ is used to compute $p(\bx_{t} | \bx_{0:t-1}, \by_{1:t-1}, \btheta)$ in Eq. \eqref{eq:Multi transitionx 2nd}, that is also necessary to calculate the predictive density $p(\bx_{0:t} | \by_{1:t-1},\btheta)$ of the second layer in Eq. \eqref{eq:Multi predx 2nd}. 
In the same vein, the predictive density $p(\bz_{h(t-1)+1:ht} | \bx_{0:t-1}, \by_{1:t-1}, \btheta)$ is used to compute the \gls{pdf} $p(\bz_{h(t-1)+1:ht}| \bx_{0:t}, \by_{1:t-1}, \btheta)$ in Eq. \eqref{eq:Multi predz weird 2nd}, that is used, in turn, to obtain the likelihood $p(\by_{t} | \bx_{0:t}, \by_{1:t-1},\btheta)$ in Eq. \eqref{eq:Multi likelihood 2nd}.

\begin{figure}[!ht]
	\centering
    {\hspace*{-1cm}
    \begin{tikzpicture}[scale=0.88]

\node (hmss) at (3.5,14.3) {\textbf{Optimal smoother}};

\node[draw, rectangle,align=center] (timeprev) at (-0.9,10.4) {$t-1$};
\node[draw, rectangle,align=center] (time) at (-1.2,8.5) {$t$};
\node[draw, rectangle,align=center] (time) at (-0.9,1.7) {$t+1$};

\node[draw,rectangle,align=center] (obs1) at (12.9,9.5) {$\by_{t-1}$};
\node[draw,rectangle,align=center] (obs2) at (12.9,5) {$\by_t$};

\draw [thin, draw=black] (-1.5,2.25) -- (-1.5,8.75) -- (12.2,8.75) -- (12.2,2.25) -- cycle;
\draw [thin, draw=black] (-1.5,9) -- (-1.5,10.7) -- (12.2,10.7) -- (12.2,9) -- cycle;
\draw [thin, draw=black] (-1.5,1.5) -- (-1.5,2) -- (12.2,2) -- (12.2,1.5);
\draw [dashed] (-1.5,1.5) -- (-1.5,1);
\draw [dashed] (12.2,1.5) -- (12.2,1);

\node[text width=5.0cm,align=center] (jointpost) at (0.25,13.8) {1$^{st}$ layer};
\node[draw, rectangle, text width=2.8cm,align=center, fill=white] (init) at (0.25,12.8) {Initialization $p(\btheta)$, $p(\bx_0)$, $p(\bz_0)$};
\coordinate (aux) at (0.25,12);
\node[draw, rectangle,text width=2cm,align=center] (prevpost1) at (0.25,9.5) {$p(\btheta|\by_{1:t-1})$};
\node[draw, rectangle, text width=2.5cm,align=center, fill=white] (lik1) at (0.25,5) {Likelihood $p(\by_t|\by_{1:t-1},\btheta)$};
\node[draw, rectangle, text width=2.5cm,align=center, fill=white] (post1) at (0.25,3) {Posterior \gls{pdf} $p(\btheta|\by_{1:t})$};

\node[text width=5.0cm,align=center] (jointpost) at (3.5,13.8) {2$^{nd}$ layer};
\node[draw, rectangle,text width=2.9cm,align=center] (prevpost2) at (3.5,9.5) {$p(\bx_{0:t-1}|\by_{1:t-1},\btheta)$};
\node[draw, rectangle, text width=2.5cm,align=center, fill=white] (pred2) at (3.5,7.6) {Predictive \gls{pdf} $p(\bx_{0:t}|\by_{1:t-1},\btheta)$};
\node[draw, rectangle, text width=3cm,align=center, fill=white] (lik2) at (3.5,5) {Likelihood $p(\by_t|\bx_{0:t},\by_{1:t-1},\btheta)$};
\node[draw, rectangle, text width=2.5cm,align=center, fill=white] (post2) at (3.5,3) {Posterior \gls{pdf} $p(\bx_{0:t}|\by_{1:t},\btheta)$};

\node[text width=5.0cm,align=center] (jointpost) at (8.7,13.8) {3$^{rd}$ layer};
\node[draw, rectangle,text width=5.45cm,align=center] (prevpost3) at (8.7,9.5) {$p(\bz_{h(t-2)+1:h(t-1)}|\bx_{0:t-1},\by_{1:t-1},\btheta)$};
\node[draw, rectangle, text width=4.85cm,align=center, fill=white] (pred3) at (8.7,7.6) {Predictive \glspl{pdf} $p(\bz_{h(t-1)+1:ht}|\bx_{0:t-1},\by_{1:t-1},\btheta)$ \\$p(\bz_{h(t-1)+1:ht}|\bx_{0:t},\by_{1:t-1},\btheta)$};
\node[draw, rectangle, text width=2.5cm,align=center, fill=white] (lik3) at (8.7,5) {Likelihood $p(\by_t|\bz_{ht},\bx_t,\btheta)$};
\node[draw, rectangle, text width=4.2cm,align=center, fill=white] (post3) at (8.7,3) {Posterior \gls{pdf} $p(\bz_{h(t-1)+1:ht}|\bx_{0:t},\by_{1:t},\btheta)$};

\draw [thick] (init) -- (aux);
\draw [thick] (aux) -| (0.25,11.75);
\draw [thick] (aux) -| (3.5,11.75) ;
\draw [thick] (aux) -| (8.7,11.75) ;
\draw [thick,dashed] (0.25,11.75) -- (0.25,11);
\draw [thick,dashed] (3.5,11.75)  -| (3.5,11) ;
\draw [thick,dashed] (8.7,11.75)  -| (8.7,11) ;
\draw [thick,->] (0.25,11) -- (prevpost1);
\draw [thick,->] (3.5,11) -- (prevpost2);
\draw [thick,->] (8.7,11) -- (prevpost3);

\draw [thick,->] (prevpost1) -- (lik1);
\draw [thick,->] (lik1) -- (post1);
\draw [thick,->] (prevpost2) -- (pred2);
\draw [thick,->] (pred2) -- (lik2);
\draw [thick,->] (lik2) -- (post2);
\draw [thick,->] (prevpost3) -- (pred3);
\draw [thick,->] (pred3) -- (lik3);
\draw [thick,->] (lik3) -- (post3);

\draw [thick,->] (post1) -- (0.25,1.75);
\draw [thick,dashed] (0.25,2) -- (0.25,1.25);
\draw [thick,->] (post2) -- (3.5,1.75);
\draw [thick,dashed] (3.5,1.75) -- (3.5,1.25);
\draw [thick,->] (post3) -- (8.7,1.75);
\draw [thick,dashed] (8.7,1.75) -- (8.7,1.25);

\draw [thick,->] (obs2) -- (lik3);
\draw [thick,->] (obs1) -- (prevpost3);

\draw [dashed,->] (pred2) -- (lik1);
\draw [dashed,->] (lik2) -- (lik1);
\draw [dashed,->] (pred3) -- (pred2);
\draw [dashed,->] (pred3) -- (lik2);
\draw [dashed,->] (lik3) -- (lik2);

\end{tikzpicture}}
    \caption{Schematic depiction of the optimal smoother. Each column represents a layer of computation and the dependencies among \glspl{pdf} are indicated by arrows. The dashed arrows are used to show relations among different layers while the solid arrows represent dependencies in the same layer.} 
\label{figscheme}
\end{figure}

%
\section{Approximate smoothing algorithms} \label{sec:Multi MultiscaleNestedFilter}

\subsection{Outline}

The optimal algorithm described in Section \ref{sec:Multi Optimalfilter} cannot be implemented exactly for most practical models. Instead, one needs to devise suitable approximations that can be implemented numerically in an efficient way. One possible approach is a full-blown \gls{smc} implementation that extends the NPF of \cite{Crisan18bernoulli}. However, such a scheme with three layers of computation results in a prohibitive computational cost. Instead, we introduce herein two different algorithms that combine \gls{smc} and Gaussian approximations at the different layers. The resulting algorithms can be implemented numerically in a more efficient manner and are suitable for parallelization, which leads to very fast run-times.

Both methods describe a numerical approximate smoother where the probability measure $p(\btheta | \by_{1:t})d\btheta$ is approximated using \gls{smc}, i.e., both schemes implement \gls{smc} in the first layer.
However, we implement different methods for the approximation of the probability measure $p(\bx_{0:t} | \by_{1:t}, \btheta)d\bx_{0:t}$ and the conditional smoothing \gls{pdf} $p(\bz_{h(t-1)+1:ht} | \bx_{0:t}, \by_{1:t}, \btheta)$. 
In the first scheme, we use another \gls{smc} at the second layer, together with a bank of \glspl{ukf} \cite{Julier04,Menegaz15} to approximate (as Gaussian) the conditional densities to be computed at the third layer. This implementation has great potential for parallelization, but it is computationally costly nevertheless. Hence, we also introduce a second, less demanding scheme that utilizes the same \gls{smc} scheme at the first layer but employs \glspl{enkf} \cite{Evensen03} at the second layer, and simple \glspl{ekf} \cite{Anderson79} to approximate the densities needed in the third layer, respectively. A numerical study of performance is carried out in Section \ref{sec:Multi Example} for both implementations.


%


\subsection{SMC implementation of layer 1}

Algorithm \ref{alg:Multi FirstScheme SMC1} describes the first layer of the nested smoother for both schemes, which aims at the approximation of the posterior distribution of the parameters. It receives as input the prior \glspl{pdf} of the parameters, $p(\btheta)$, and the states, $p(\bx_0)$ and $p(\bz_0)$. In the initialization step, $p(\btheta)$ is used to generate $N$  \gls{iid} Monte Carlo parameter samples $\{\btheta_0^{(i)}\}_{i=1}^N$. The prior \glspl{pdf} of the states are used in the initialization steps of layers 2 and 3. In both schemes, the \gls{pdf} $p(\bx_0)$ is used to initialize the slow state in layer 2 by generating a set of initial samples $\{\bx_0^{(i,j)}\}_{j=1}^J$ for each parameter sample, $\btheta_0^{(i)}$. 
The initialization of the fast states differs between both schemes since we use $p(\bz_0)$ to generate a set of points $\{\bz_0^{(i,j,l)}\}_{l=0}^L$ in the \gls{ukf}, and mean and covariance matrix, $\widehat{\bz}_0^{(i,j)}$ and $\bC_0^{(i,j)}$, in the \gls{ekf}. Note that for the first scheme may still have a prohibitive computational cost, since it
generates a total of $N \times J \times L$  particles in the joint space of the parameters, the slow states and the fast states (if we count sigma-points as deterministic particles).
The second scheme describes a computationally-lighter procedure.  

Additionally, a Markov kernel $\kappa_N^{\btheta'}(d\btheta)$ is needed for the jittering of parameter samples \cite{Crisan18bernoulli}, i.e., to draw a new set of particles, $\{\widetilde{\btheta}_{t}^{(i)}\}_{i=1}^N$, at each discrete-time step. 
 This is needed to preserve the diversity of the particles. Otherwise after a few resampling steps the parameter particles would be reduced to just a few distinct values and the filter would collapse. This technique also allows us to keep the scheme running recursively, i.e, without reprocessing the whole sequence of observations from scratch.

\vspace{2mm}

\par\noindent\rule{0.9\textwidth}{0.4pt}
\begin{algoritmo} {\gls{smc} approximation of $p(\btheta | \by_{1:t})$ in the first method \label{alg:Multi FirstScheme SMC1}}
		
		\textbf{Inputs}
		\begin{itemize}
			\item[-] Prior distributions $p(\btheta)$, $p(\bx_0)$ and $p(\bz_0)$.
			\item[-] A Markov kernel $\kappa_N^{\btheta'}(d\btheta)$ which, given $\btheta'$, generates jittered parameters $\btheta \in \Reals^{d_\theta}$.
		\end{itemize}
		
		\textbf{Initialization:} 
		\begin{itemize}
			\item[-] Draw $N$ i.i.d. sample ${\btheta}_0^{(i)} \sim p(\btheta)$, for $i = 1,\ldots,N$.
			
		\end{itemize}
		
		\textbf{Procedure} For $t > 0$:
		\begin{enumerate}

            \item Jittering: draw $N$ i.i.d samples $\widetilde{\btheta}_{t}^{(i)}$ from $\kappa_N^{\btheta_{t-1}^{(i)}}(d\btheta)$.
			
			\item For each $i=1,\ldots,N$, run algorithm of layer 2 to:
			
			\begin{enumerate}
               \item For each $j=1,\ldots,J$, run algorithm of layer 3 to:
               \begin{enumerate}
                   
                   \item Evaluate the likelihood $p(\by_t | \bz_{ht}, \bx_t^{(i,j)}, \widetilde{\btheta}_t^{(i)})$ and  compute the approximate marginal likelihood $\widetilde{u}_{t}^{(i,j)} = \widehat{p} (\by_{t} | \bx_{0:t}^{(i,j)}, \by_{1:t-1}, \widetilde{\btheta}_{t}^{(i)})$. \label{stepal1likelihhod3}
                   \item Approximate the posterior distribution $\widehat{p}(\bz_{h(t-1)+1:ht} | \bx_{0:t}^{(i,j)}, \by_{1:t}, \widetilde{\btheta}_t^{(i)})$.
                \end{enumerate}


                \item Compute the approximate marginal likelihood $\widetilde{v}_{t}^{(i)} = \widehat{p} (\by_{t} | \by_{1:t-1}, \widetilde{\btheta}_{t}^{(i)})$. \label{stepal1likelihhod2}
				\item Approximate the posterior \gls{pdf} $\widehat{p}( \bx_{0:t}| \by_{1:t},\widetilde{\btheta}_t^{(i)})$.
				
			\end{enumerate}

            \item Normalize the weights \label{stepal1normweights}
				\begin{equation}
				\quad v_{t}^{(i)} = \frac{\widetilde{v}_{t}^{(i)}}{\sum_{k=1}^{N} \widetilde{v}_{t}^{(k)}}. \label{eqal1normweights}
				\end{equation} 

			\item Resample: draw indices $m_1, \ldots, m_N$ with probability $v_{t}^{(1)}, \ldots, v_{t}^{(N)}$. For $i=1,\ldots,N$ set \label{stepal1resampling}
			\begin{eqnarray}
                \btheta_{t}^{(i)} &=& \widetilde{\btheta}_{t}^{(m_i)}, \nonumber \\
                \widehat{p}(\bx_{0:t} |\by_{1:t},{\btheta}_t^{(i)}) &=& \widehat{p}(\bx_{0:t} | \by_{1:t},\widetilde{\btheta}_t^{(m_i)}) \quad \text{and}\nonumber \\
                \widehat{p}(\bz_{h(t-1)+1:ht} | \bx_{0:t}^{(i,j)}, \by_{1:t}, {\btheta}_t^{(i)}) &=& \widehat{p}(\bz_{h(t-1)+1:ht} | \bx_{0:t}^{(i,j)}, \by_{1:t}, \widetilde{\btheta}_t^{(m_i)}), \nonumber
			\end{eqnarray}
            for all $j=1,\ldots,J$.

		\end{enumerate}
		
		\textbf{Outputs:} $\{\btheta_{t}^{(i)}, \widehat{p}(\bx_{0:t} | \by_{1:t},{\btheta}_t^{(i)}), \{ \widehat{p}(\bz_{h(t-1)+1:ht} | \bx_{0:t}^{(i,j)}, \by_{1:t}, {\btheta}_t^{(i)}) \}_{j=1}^J  \}_{i=1}^N$. 
		
\end{algoritmo}
\par\noindent\rule{0.9\textwidth}{0.4pt}

\newpage
At every time step $t$ (in the slow time scale), we run layers 2 and 3 of the scheme.
To be more specific, for each $j$-th sample of the slow states, we run the algorithm of layer 3 in order to evaluate the likelihood $p(\by_t | \bz_{ht}, \bx_t^{(i,j)}, \widetilde{\btheta}_t^{(i)})$ and to approximate the marginal likelihood, $\widetilde{u}_t^{(i,j)}$, of Eq. \eqref{eq:Multi likelihood 2nd}. In this layer, we also approximate the posterior \gls{pdf} of the fast states, $p(\bz_{h(t-1)+1:ht}| \bx_{0:t}^{(i,j)}, \by_{1:t}, \widetilde{\btheta}_t^{(i)} )$.
Once layer 3 is done, in layer 2 we compute the approximate likelihood for each particle $\widetilde{\btheta}_{t}^{(i)}$, namely 
$$
\widehat{p} (\by_{t} | \by_{1:t-1}, \widetilde{\btheta}_{t}^{(i)}) \approx p(\by_{t} | \by_{1:t-1}, \widetilde{\btheta}_{t}^{(i)}),
$$ 
in order to obtain the non-normalized weights $\{\widetilde{v}_{t}^{(i)}\}_{i=1}^N$. The posterior distribution of the slow states, $\widehat{p}( \bx_{0:t} | \by_{1:t},\widetilde{\btheta}_t^{(i)})$ , is also approximated here.
%
Finally, in layer 1, we normalize the weights in order to resample not only the parameter particles $\widetilde{\btheta}_{t}^{(i)}$, but also their associated approximations of the distributions of the state variables. Full details on the implementation of the two approximate schemes are provided in \ref{ap:1} and \ref{ap:2}.

\section{Example} \label{sec:Multi Example}

\subsection{Stochastic two-scale Lorenz 96 model} \label{subsec:Multi StochasticL96model}

In order to illustrate the application of the methods described in Section \ref{sec:Multi MultiscaleNestedFilter}, we consider a stochastic version of the two-scale Lorenz 96 model \cite{Arnold13}, which depends on a set of fixed parameters, a set of fast variables and a set of slow variables. The slow variables are represented by a $d_x$-dimensional vector, $\bx$, while the fast variables, $\bz$, are $d_z$-dimensional. {Let us assume there are $R$ fast variables per slow variable, therefore $d_z = R d_x$.} The system is described, in continuous-time $\tau$, by the \glspl{sde} 
\begin{eqnarray}
{d x_{j}}&= \bigg[ -x_{j-1}(x_{j-2}-x_{j+1}) - x_j+F - \frac{H C}{B} \sum_{l=(j-1)R}^{Rj-1} z_l \bigg] d \tau+ \sigma_x d v_j , \label{eqlorenz96sdex}
\\
{d z_{l}}&= \bigg[ -CBz_{l+1}(z_{l+2}-z_{l-1}) - Cz_l + \frac{CF}{B} +\frac{HC}{B}x_{{\lfloor (l-1)/R \rfloor}} \bigg] d \tau + \sigma_z d w_l, \label{eqlorenz96sdez}
\end{eqnarray}
where $j = 0,\ldots,d_x-1$, $l = 0,\ldots, d_z-1$; $\bv=(v_0,\ldots,v_{d_x-1})^\top$ and $\bw=(w_0,\ldots,w_{d_z-1})^\top$ are, respectively, $d_x$- and $d_z$-dimensional vectors of independent standard Wiener processes; $\sigma_x > 0$ and $\sigma_z > 0$ are known scale parameters and $\btheta=(F,H,C,B)^\top \in \Reals$ are static model parameters. Using a micro-macro solver \cite{Vanden03,Weinan05} that runs an Euler-Maruyama scheme at each time-scale to integrate Eqs. \eqref{eqlorenz96sdex}--\eqref{eqlorenz96sdez}, the discrete-time state equation can be written as
\begin{eqnarray}
x_{t+1,j} &=& x_{t,j} + \Delta_x(f_{\bx,j}(\bx_{t}, \btheta) + g_{\bx,j}(\overline{\bz}_{t+1},\btheta) ) + \sqrt{\Delta_x} \sigma_x v_{t+1,j}, \label{eqLorenz96X}
\\
z_{n+1,l} &=& z_{n,l} + \Delta_z( f_{\bz,l}(\bx_{\lfloor\frac{n}{h}\rfloor}, \btheta) + g_{\bz,l}(\bz_n, \btheta)) + \sqrt{\Delta_z} \sigma_z w_{n+1,l}, \label{eqLorenz96Z}
\end{eqnarray}
where 
$$
\bx_{t}=(x_{t,0}, \ldots, x_{t,d_x-1})^\top \quad \text{and} \quad
{\bz_n=(z_{n,0}, \ldots, z_{n,d_z-1})^\top}
$$
are the discrete-time slow and fast variables, respectively; $\overline{\bz}_{t}$ is the time-average
$$
\overline{\bz}_{t} = {\frac{1}{h} \sum_{n=h(t-1)+1}^{ht} \bz_n}
$$
and we denote {$\overline{\bz}_{t}=(\overline{z}_{t,0}, \ldots, \overline{z}_{t,d_z-1})^\top$; }the terms $v_{t,j}$ and $w_{n,l}$ are independent Gaussian variables with identical $\mN(\cdot|0,1)$ \gls{pdf} for all $t$, $j$, $n$ and $l$, and the functions
\begin{eqnarray}
f_{\bx,j} &:& \Reals^{d_x} \times \Reals^{d_{\theta}} \to \Reals^{d_x}~ ~\text{and}~ ~g_{\bx,j} : \Reals^{d_z} \times \Reals^{d_{\theta}} \to \Reals^{d_x},~ ~\text{and}\nonumber\\
f_{\bz,l} &:& \Reals^{d_x} \times \Reals^{d_{\theta}} \to \Reals^{d_z} ~ ~\text{and}~ ~g_{\bz,l} : \Reals^{d_z} \times \Reals^{d_{\theta}} \to \Reals^{d_z}\nonumber
\end{eqnarray}
can be expressed as
\begin{eqnarray}
f_{\bx,j}(\bx_{t},\btheta) &=& -x_{t,j-1}(x_{t,j-2}-x_{t,j+1}) - x_{t,j} + F, \nonumber \\
g_{\bx,j}(\overline{\bz}_{t}, \btheta) &=& - \frac{H C}{B} \sum_{l=(j-1)R}^{Rj-1} \overline{z}_{t,l}, \nonumber \\
f_{\bz,l}(\bx_{t}, \btheta) &=& \frac{HC}{B}x_{t,{\lfloor (l-1)/R \rfloor}} \quad \text{and} \nonumber \\
g_{\bz,l}(\bz_n,\btheta) &=& -CBz_{n,l+1}(z_{n,l+2}-z_{n,l-1}) - Cz_{n,l} + \frac{CF}{B}. \nonumber
\end{eqnarray}

We assume that the observations are linear and Gaussian, namely,
\begin{equation}
	\by_{t} = \boldsymbol{A}_{t}
	\begin{bmatrix}
	\bx_{t} \\ 
	\bz_{ht}  
	\end{bmatrix}
	+ \br_{t}, \label{eqObs96}
\end{equation}
where $\boldsymbol{A}_{t}$ is a known $d_y \times (d_x+d_z)$ matrix and $\br_{t}$ is a $d_y$-dimensional Gaussian random vector with known covariance matrix
\begin{equation}
\bR =
\begin{bmatrix}
\sigma_{y,x}^2 \boldsymbol{I}_{d_{x}} & \boldsymbol{0} \\ 
\boldsymbol{0} & \sigma_{y,z}^2 \boldsymbol{I}_{d_{z}} 
\end{bmatrix}, \label{eqR}
\end{equation}
and $\sigma_{y,x}^2, \sigma_{y,z}^2 >0$ are known variances.

\subsection{Numerical results} \label{subsec:Multi Numericalresults}

We have run simulations for the two-scale Lorenz 96 model of Section \ref{subsec:Multi StochasticL96model}, with dimensions $d_x = 10$ and $d_z=50$. The time steps for the Euler-Maruyama integrators are $\Delta_x = 10^{-3}$ and $\Delta_z = 10^{-4}$ continuous-time units. We set the fixed parameters as $F=8$, $H=0.75$, $C=10$ and $B=15$. In order to obtain the initial states $\widehat \bx_0$ and $\widehat \bz_0$, we simulate a deterministic version of Eqs. \eqref{eqLorenz96X}--\eqref{eqLorenz96Z} ($\sigma_x = \sigma_z = 0$) for $20$ continuous-time units. We set the initial states as the values of variables $\bx$ and $\bz$ at the last time step of this simulation. This initialization is used in all simulations of this computer experiment in order to generate both ``ground truth" sequences of $\bx_{t}$ and $\bz_n$ and the associated sequences of observations $\by_{t}$. {We set the matrix $\boldsymbol{A}_{t}=\bI_{d_y}$, for $d_y = d_x + d_z$. }

In the experiments, we compare the performance of the two approximate methods (with a common layer 1 described in Section \ref{sec:Multi MultiscaleNestedFilter} and additional details provided in \ref{ap:1} and \ref{ap:2}). 
We experiment with different number of samples $N$ and $J$ in the first and second layers of the former methods. 
Additionally, for the first method (\gls{smc}-\gls{smc}-\gls{ukf}) we run the multi-scale hybrid filter with $L = 2 d_z + 1 = 101$ sigma-points for the \gls{ukf} in the third layer. We need to estimate $\btheta = [F, C, H, B]^\top$ (hence, $d_\theta = 4$). The prior for the unknown parameters is uniform, namely $p(\btheta) = \mathcal{U}([2,20]^2)$, while the priors used in the filtering algorithm for both unknown state variables are Gaussian, namely $p(\bx_0) = \mathcal{N}(\bx_0 | \widehat \bx_0, 0.1 \boldsymbol{I}_{d_x})$ and $p(\bz_0) = \mathcal{N}(\bz_0 | \widehat \bz_0, 10 \boldsymbol{I}_{d_z})$. The noise scaling factors, $\sigma_x = \frac{1}{2}$, $\sigma_z = \frac{1}{16}$, $\sigma_{y,x} = 10^{-1}$ and $\sigma_{y,z} = 10^{-3}$, are known. The jittering kernel is $\kappa_N^{\btheta'}(d\btheta) = \mathcal{N}(\btheta | \btheta', \widetilde{\sigma}^2 \boldsymbol{I}_{d_{\theta}})$, where $\widetilde{\sigma}^2 = \frac{0.05}{\sqrt{N^3}}$ {is selected following \cite{Crisan18bernoulli}}.

We assess the accuracy of the algorithms in terms of the \gls{nmse} of the estimators of the parameters, the slow state variables and the fast state variables. In the plots, we show the \glspl{nmse} computed at time $t$,
\begin{eqnarray}
\text{NMSE}_{\btheta,t} &=& \frac{\| \btheta_{t} - \widehat{\btheta}_{t} \|^2}{\| \btheta_{t} \|^2}, \\
\text{NMSE}_{\bx,t} &=& \frac{\| \bx_{t} - \widehat{\bx}_{t} \|^2}{\| \bx_{t} \|^2}, \\
\text{NMSE}_{\bz,t} &=& \frac{\| \bz_{ht} - \widehat{\bz}_{ht} \|^2}{ \| \bz_{ht} \|^2},
\end{eqnarray}
averaged over 50 independent simulation runs of 20 continuous-time units each, where the estimators take the form
\begin{eqnarray}
\widehat{\btheta}_{t} &=& \sum_{i=1}^{N} v_{t}^{(i)} {\btheta}^{(i)}_t, \\
\widehat{\bx}_{t} &=& \sum_{i=1}^{N} \sum_{j=1}^{J} v_{t}^{(i)} u_{t}^{(i,j)} {\bx}^{(i,j)}_{t} \quad \text{and} \\
\widehat{\bz}_{ht} &=& {\sum_{i=1}^{N} \sum_{j=1}^{J} \sum_{l=0}^{L} v_{t}^{(i)} u_{t}^{(i,j)} \lambda_{ht}^{(i,j,l)} {\bz}^{(i,j,l)}_{ht|ht}, }
\end{eqnarray}
for the first method. Here, the set of particles and weights $\{ \btheta_t^{(i)},v_t^{(i)} \}_{i=1}^N$ is the output of the \gls{smc} of the first layer, the set $\{ \bx_t^{(i,j)},u_t^{(i,j)} \}_{j=1}^J$, for $i=1,\ldots,N$, is the output of the \gls{smc} of the second layer, and the set of points and weights $\{ \bz_{ht|ht}^{(i,j,l)},\lambda_{ht}^{(i,j,l)} \}_{l=0}^{L-1}$, for $i=1,\ldots,N$ and $j=1,\ldots,J$, is the output of the \gls{ukf} of the third layer of computation.
See Appendix \ref{ap:1} for details on the computation of these sets.
For the second method the estimators of the state variables are 
\begin{eqnarray}
\widehat{\bx}_{t} &=&  \frac{1}{J} \sum_{i=1}^{N} \sum_{j=1}^{J} v_{t}^{(i)} {\bx}^{(i,j)}_{t} \quad \text{and} \\
\widehat{\bz}_{ht} &=& \frac{1}{J} \sum_{i=1}^{N} \sum_{j=1}^{J} v_{t}^{(i)} \widehat{\bz}^{(i,j)}_{ht|ht},
\end{eqnarray}
where $\bx_{t}^{(i,j)}$ is the $j$-th member of the ensemble $\bX_{t|t}^{(i)}$ in the second layer of the SMC-EnKF-EKF algorithm, and $\widehat{\bz}_{ht|ht}^{(i,j)}$ is the updated mean in the third layer of computation (i.e., in the \gls{ekf}).
Details on how to compute these approximations are in Appendix \ref{ap:2}.

Figure \ref{fig:Multi changeJ} shows the performance of the proposed methods for different values of $J$ (number of samples in the second layer) and $N=20$. This is evaluated in terms of averaged \gls{nmse}$_{\btheta}$, \gls{nmse}$_{\bx}$ and \gls{nmse}$_{\bz}$ together with the run time. These simulations have been carried in MatLab R2020a on a Linux machine with an Intel(R) Xeon(R) Silver 4210 CPU @ 2.20GHz processor and 20 cores. The first method (\gls{smc}-\gls{smc}-\gls{ukf}) shows an improvement in the accuracy as the number of samples $J$ increases, although this improvement is only significant for the slow state (Fig. \ref{figNMSExvsJ}). The second method (\gls{smc}-\gls{enkf}-\gls{ekf}) remains stable with $J$. The second method outperforms the first one in accuracy of the parameter estimation (Fig. \ref{figNMSEthetavsJ}) as well as the slow state estimation (Fig. \ref{figNMSExvsJ}). However, the first method obtains a better \gls{nmse}$_{\bz}$. Additionally, the second method runs faster since the computational cost is considerably lower.


\begin{figure}[t]
	\centering
	\begin{subfigure}[htb]{0.47\linewidth}
		\centering
%
%

\definecolor{mycolor1}{rgb}{1.00000,0.00000,1.00000}%
\begin{tikzpicture}

\begin{axis}[%
width=\wfigtwo,
height=\hfigtwo,
at={(1.011in,0.642in)},
scale only axis,
unbounded coords=jump,
xmin=0.9,
xmax=4.1,
xtick={1.4,2.4,3.4},
xticklabels={{20},{50},{100}},
xlabel={J},
xmajorgrids,
ymin=0.001,
ymax=1,
ylabel={NMSE$_{\btheta}$},
ymajorgrids,
yminorgrids,
ymode=log,
axis background/.style={fill=white},
legend style={legend cell align=left,align=left,draw=white!15!black}
]
\addplot [color=blue,dashed,line width=0.6pt,forget plot]
  table[row sep=crcr]{%
1.3	0.158533202305466\\
1.3	0.199599256968954\\
};
\addplot [color=blue,dashed,line width=0.6pt,forget plot]
  table[row sep=crcr]{%
2.3	0.096098582722346\\
2.3	0.104079773532239\\
};
\addplot [color=blue,dashed,line width=0.6pt,forget plot]
  table[row sep=crcr]{%
3.3	0.0920824473865342\\
3.3	0.179139585957066\\
};
\addplot [color=blue,dashed,line width=0.6pt,forget plot]
  table[row sep=crcr]{%
1.3	0.0101692235225769\\
1.3	0.0318441233184488\\
};
\addplot [color=blue,dashed,line width=0.6pt,forget plot]
  table[row sep=crcr]{%
2.3	0.00431769680413072\\
2.3	0.0145152532295724\\
};
\addplot [color=blue,dashed,line width=0.6pt,forget plot]
  table[row sep=crcr]{%
3.3	0.00475156267628708\\
3.3	0.0197146972329995\\
};
\addplot [color=blue,solid,line width=0.6pt,forget plot]
  table[row sep=crcr]{%
1.255	0.199599256968954\\
1.345	0.199599256968954\\
};
\addplot [color=blue,solid,line width=0.6pt,forget plot]
  table[row sep=crcr]{%
2.255	0.104079773532239\\
2.345	0.104079773532239\\
};
\addplot [color=blue,solid,line width=0.6pt,forget plot]
  table[row sep=crcr]{%
3.255	0.179139585957066\\
3.345	0.179139585957066\\
};
\addplot [color=blue,solid,line width=0.6pt,forget plot]
  table[row sep=crcr]{%
1.255	0.0101692235225769\\
1.345	0.0101692235225769\\
};
\addplot [color=blue,solid,line width=0.6pt,forget plot]
  table[row sep=crcr]{%
2.255	0.00431769680413072\\
2.345	0.00431769680413072\\
};
\addplot [color=blue,solid,line width=0.6pt,forget plot]
  table[row sep=crcr]{%
3.255	0.00475156267628708\\
3.345	0.00475156267628708\\
};
\addplot [color=blue,line width=0.6pt,solid]
  table[row sep=crcr]{%
1.21	0.0318441233184488\\
1.21	0.158533202305466\\
1.39	0.158533202305466\\
1.39	0.0318441233184488\\
1.21	0.0318441233184488\\
};

\addplot [color=blue,solid,line width=0.6pt,forget plot]
  table[row sep=crcr]{%
2.21	0.0145152532295724\\
2.21	0.096098582722346\\
2.39	0.096098582722346\\
2.39	0.0145152532295724\\
2.21	0.0145152532295724\\
};
\addplot [color=blue,solid,line width=0.6pt,forget plot]
  table[row sep=crcr]{%
3.21	0.0197146972329995\\
3.21	0.0920824473865342\\
3.39	0.0920824473865342\\
3.39	0.0197146972329995\\
3.21	0.0197146972329995\\
};
\addplot [color=blue,solid,line width=0.6pt,forget plot]
  table[row sep=crcr]{%
1.21	0.0683208250964539\\
1.39	0.0683208250964539\\
};
\addplot [color=blue,solid,line width=0.6pt,forget plot]
  table[row sep=crcr]{%
2.21	0.031726634365873\\
2.39	0.031726634365873\\
};
\addplot [color=blue,solid,line width=0.6pt,forget plot]
  table[row sep=crcr]{%
3.21	0.0539390725589018\\
3.39	0.0539390725589018\\
};
\addplot [color=black,only marks,mark=+,mark options={solid,draw=red},line width=0.6pt,forget plot]
  table[row sep=crcr]{%
1.3	0.52293850428406\\
1.3	1.10901317842438\\
};
\addplot [color=black,only marks,mark=+,mark options={solid,draw=red},line width=0.6pt,forget plot]
  table[row sep=crcr]{%
2.3	0.232834416212537\\
2.3	0.257842210738612\\
2.3	0.503211500316656\\
2.3	0.864641929187627\\
};
\addplot [color=black,only marks,mark=+,mark options={solid,draw=red},line width=0.6pt,forget plot]
  table[row sep=crcr]{%
3.3	0.328985528762008\\
3.3	0.392768156944399\\
3.3	0.687262989392093\\
};
\addplot [color=red,dashed,line width=0.6pt,forget plot]
  table[row sep=crcr]{%
1.5	0.0736633644277398\\
1.5	0.108827301255656\\
};
\addplot [color=red,dashed,line width=0.6pt,forget plot]
  table[row sep=crcr]{%
2.5	0.0357484666438214\\
2.5	0.0390621671798566\\
};
\addplot [color=red,dashed,line width=0.6pt,forget plot]
  table[row sep=crcr]{%
3.5	0.0408895765040056\\
3.5	0.0737861906480782\\
};
\addplot [color=red,dashed,line width=0.6pt,forget plot]
  table[row sep=crcr]{%
1.5	0.00551145297983492\\
1.5	0.0151525304967675\\
};
\addplot [color=red,dashed,line width=0.6pt,forget plot]
  table[row sep=crcr]{%
2.5	0.00125795354400581\\
2.5	0.00633946535691591\\
};
\addplot [color=red,dashed,line width=0.6pt,forget plot]
  table[row sep=crcr]{%
3.5	0.00150026705500902\\
3.5	0.00817543007743051\\
};
\addplot [color=red,solid,line width=0.6pt,forget plot]
  table[row sep=crcr]{%
1.455	0.108827301255656\\
1.545	0.108827301255656\\
};
\addplot [color=red,solid,line width=0.6pt,forget plot]
  table[row sep=crcr]{%
2.455	0.0390621671798566\\
2.545	0.0390621671798566\\
};
\addplot [color=red,solid,line width=0.6pt,forget plot]
  table[row sep=crcr]{%
3.455	0.0737861906480782\\
3.545	0.0737861906480782\\
};
\addplot [color=red,solid,line width=0.6pt,forget plot]
  table[row sep=crcr]{%
1.455	0.00551145297983492\\
1.545	0.00551145297983492\\
};
\addplot [color=red,solid,line width=0.6pt,forget plot]
  table[row sep=crcr]{%
2.455	0.00125795354400581\\
2.545	0.00125795354400581\\
};
\addplot [color=red,solid,line width=0.6pt,forget plot]
  table[row sep=crcr]{%
3.455	0.00150026705500902\\
3.545	0.00150026705500902\\
};
\addplot [color=red,solid,line width=0.6pt,forget plot]
  table[row sep=crcr]{%
1.41	0.0151525304967675\\
1.41	0.0736633644277398\\
1.59	0.0736633644277398\\
1.59	0.0151525304967675\\
1.41	0.0151525304967675\\
};
\addplot [color=red,solid,line width=0.6pt]
  table[row sep=crcr]{%
2.41	0.00633946535691591\\
2.41	0.0357484666438214\\
2.59	0.0357484666438214\\
2.59	0.00633946535691591\\
2.41	0.00633946535691591\\
};

\addplot [color=red,solid,line width=0.6pt,forget plot]
  table[row sep=crcr]{%
3.41	0.00817543007743051\\
3.41	0.0408895765040056\\
3.59	0.0408895765040056\\
3.59	0.00817543007743051\\
3.41	0.00817543007743051\\
};
\addplot [color=red,solid,line width=0.6pt,forget plot]
  table[row sep=crcr]{%
1.41	0.0260427990702596\\
1.59	0.0260427990702596\\
};
\addplot [color=red,solid,line width=0.6pt,forget plot]
  table[row sep=crcr]{%
2.41	0.0134065598011601\\
2.59	0.0134065598011601\\
};
\addplot [color=red,solid,line width=0.6pt,forget plot]
  table[row sep=crcr]{%
3.41	0.0132038022646453\\
3.59	0.0132038022646453\\
};
\addplot [color=black,only marks,mark=+,mark options={solid,draw=red},line width=0.6pt,forget plot]
  table[row sep=crcr]{%
2.5	0.0807144560140274\\
2.5	0.0816833814116167\\
2.5	0.0922651114907573\\
2.5	0.0975604234435486\\
};

\end{axis}
\end{tikzpicture}%
		\caption{\gls{nmse}$_{\btheta}$ for different $J$ in the second layer.}
		\label{figNMSEthetavsJ}
	\end{subfigure}
    \hfill
	\begin{subfigure}[htb]{0.47\linewidth}
		\centering
%
%
\definecolor{mycolor1}{rgb}{1.00000,0.00000,1.00000}%

\begin{tikzpicture}

\begin{axis}[%
width=\wfigtwo,
height=\hfigtwo,
at={(1.011in,0.642in)},
scale only axis,
xmin=0.9,
xmax=4.1,
xtick={1.4,2.4,3.4},
xticklabels={{20},{50},{100}},
xlabel={J},
xmajorgrids,
ymin=0.0001,
ymax=0.1,
ylabel={NMSE$_{\bx}$},
ymajorgrids,
yminorgrids,
ymode=log,
axis background/.style={fill=white},
legend style={legend cell align=left,align=left,draw=white!15!black}
]
\addplot [color=blue,dashed,line width=0.6pt,forget plot]
  table[row sep=crcr]{%
1.3	0.0384689597541337\\
1.3	0.0462705846629572\\
};
\addplot [color=blue,dashed,line width=0.6pt,forget plot]
  table[row sep=crcr]{%
2.3	0.00725486529541854\\
2.3	0.011349314243044\\
};
\addplot [color=blue,dashed,line width=0.6pt,forget plot]
  table[row sep=crcr]{%
3.3	0.00665154142789136\\
3.3	0.00799834216638002\\
};
\addplot [color=blue,dashed,line width=0.6pt,forget plot]
  table[row sep=crcr]{%
1.3	0.00359966105214971\\
1.3	0.00802962049234504\\
};
\addplot [color=blue,dashed,line width=0.6pt,forget plot]
  table[row sep=crcr]{%
2.3	0.00110311516414016\\
2.3	0.00274345509212269\\
};
\addplot [color=blue,dashed,line width=0.6pt,forget plot]
  table[row sep=crcr]{%
3.3	0.00101048514700265\\
3.3	0.00274928795663302\\
};
\addplot [color=blue,solid,line width=0.6pt,forget plot]
  table[row sep=crcr]{%
1.255	0.0462705846629572\\
1.345	0.0462705846629572\\
};
\addplot [color=blue,solid,line width=0.6pt,forget plot]
  table[row sep=crcr]{%
2.255	0.011349314243044\\
2.345	0.011349314243044\\
};
\addplot [color=blue,solid,line width=0.6pt,forget plot]
  table[row sep=crcr]{%
3.255	0.00799834216638002\\
3.345	0.00799834216638002\\
};
\addplot [color=blue,solid,line width=0.6pt,forget plot]
  table[row sep=crcr]{%
1.255	0.00359966105214971\\
1.345	0.00359966105214971\\
};
\addplot [color=blue,solid,line width=0.6pt,forget plot]
  table[row sep=crcr]{%
2.255	0.00110311516414016\\
2.345	0.00110311516414016\\
};
\addplot [color=blue,solid,line width=0.6pt,forget plot]
  table[row sep=crcr]{%
3.255	0.00101048514700265\\
3.345	0.00101048514700265\\
};
\addplot [color=blue,line width=0.6pt,solid]
  table[row sep=crcr]{%
1.21	0.00802962049234504\\
1.21	0.0384689597541337\\
1.39	0.0384689597541337\\
1.39	0.00802962049234504\\
1.21	0.00802962049234504\\
};

\addplot [color=blue,solid,line width=0.6pt,forget plot]
  table[row sep=crcr]{%
2.21	0.00274345509212269\\
2.21	0.00725486529541854\\
2.39	0.00725486529541854\\
2.39	0.00274345509212269\\
2.21	0.00274345509212269\\
};
\addplot [color=blue,solid,line width=0.6pt,forget plot]
  table[row sep=crcr]{%
3.21	0.00274928795663302\\
3.21	0.00665154142789136\\
3.39	0.00665154142789136\\
3.39	0.00274928795663302\\
3.21	0.00274928795663302\\
};
\addplot [color=blue,solid,line width=0.6pt,forget plot]
  table[row sep=crcr]{%
1.21	0.0100693150598859\\
1.39	0.0100693150598859\\
};
\addplot [color=blue,solid,line width=0.6pt,forget plot]
  table[row sep=crcr]{%
2.21	0.00485974722392104\\
2.39	0.00485974722392104\\
};
\addplot [color=blue,solid,line width=0.6pt,forget plot]
  table[row sep=crcr]{%
3.21	0.0039694636480501\\
3.39	0.0039694636480501\\
};
\addplot [color=black,only marks,mark=+,mark options={solid,draw=red},line width=0.6pt,forget plot]
  table[row sep=crcr]{%
1.3	0.17340670924678\\
1.3	0.254724048546887\\
1.3	0.33\\
};
\addplot [color=black,only marks,mark=+,mark options={solid,draw=red},line width=0.6pt,forget plot]
  table[row sep=crcr]{%
2.3	0.33\\
2.3	0.33\\
};
\addplot [color=black,only marks,mark=+,mark options={solid,draw=red},line width=0.6pt,forget plot]
  table[row sep=crcr]{%
3.3	0.0386125437695603\\
3.3	0.264349920000347\\
3.3	0.33\\
};
\addplot [color=red,dashed,line width=0.6pt,forget plot]
  table[row sep=crcr]{%
1.5	0.00194320571885995\\
1.5	0.00213876173068695\\
};
\addplot [color=red,dashed,line width=0.6pt,forget plot]
  table[row sep=crcr]{%
2.5	0.00146485548468875\\
2.5	0.00239281805628899\\
};
\addplot [color=red,dashed,line width=0.6pt,forget plot]
  table[row sep=crcr]{%
3.5	0.00149594079699958\\
3.5	0.0021116547546972\\
};
\addplot [color=red,dashed,line width=0.6pt,forget plot]
  table[row sep=crcr]{%
1.5	0.000728858927399826\\
1.5	0.00115078972649986\\
};
\addplot [color=red,dashed,line width=0.6pt,forget plot]
  table[row sep=crcr]{%
2.5	0.000293204346510265\\
2.5	0.000597557886643933\\
};
\addplot [color=red,dashed,line width=0.6pt,forget plot]
  table[row sep=crcr]{%
3.5	0.00034158764553078\\
3.5	0.000781392275488568\\
};
\addplot [color=red,solid,line width=0.6pt,forget plot]
  table[row sep=crcr]{%
1.455	0.00213876173068695\\
1.545	0.00213876173068695\\
};
\addplot [color=red,solid,line width=0.6pt,forget plot]
  table[row sep=crcr]{%
2.455	0.00239281805628899\\
2.545	0.00239281805628899\\
};
\addplot [color=red,solid,line width=0.6pt,forget plot]
  table[row sep=crcr]{%
3.455	0.0021116547546972\\
3.545	0.0021116547546972\\
};
\addplot [color=red,solid,line width=0.6pt,forget plot]
  table[row sep=crcr]{%
1.455	0.000728858927399826\\
1.545	0.000728858927399826\\
};
\addplot [color=red,solid,line width=0.6pt,forget plot]
  table[row sep=crcr]{%
2.455	0.000293204346510265\\
2.545	0.000293204346510265\\
};
\addplot [color=red,solid,line width=0.6pt,forget plot]
  table[row sep=crcr]{%
3.455	0.00034158764553078\\
3.545	0.00034158764553078\\
};
\addplot [color=red,solid,line width=0.6pt,forget plot]
  table[row sep=crcr]{%
1.41	0.00115078972649986\\
1.41	0.00194320571885995\\
1.59	0.00194320571885995\\
1.59	0.00115078972649986\\
1.41	0.00115078972649986\\
};
\addplot [color=red,line width=0.6pt,solid]
  table[row sep=crcr]{%
2.41	0.000597557886643933\\
2.41	0.00146485548468875\\
2.59	0.00146485548468875\\
2.59	0.000597557886643933\\
2.41	0.000597557886643933\\
};

\addplot [color=red,solid,line width=0.6pt,forget plot]
  table[row sep=crcr]{%
3.41	0.000781392275488568\\
3.41	0.00149594079699958\\
3.59	0.00149594079699958\\
3.59	0.000781392275488568\\
3.41	0.000781392275488568\\
};
\addplot [color=red,solid,line width=0.6pt,forget plot]
  table[row sep=crcr]{%
1.41	0.00171255210995195\\
1.59	0.00171255210995195\\
};
\addplot [color=red,solid,line width=0.6pt,forget plot]
  table[row sep=crcr]{%
2.41	0.00103802194880769\\
2.59	0.00103802194880769\\
};
\addplot [color=red,solid,line width=0.6pt,forget plot]
  table[row sep=crcr]{%
3.41	0.00104616567148862\\
3.59	0.00104616567148862\\
};
\addplot [color=black,only marks,mark=+,mark options={solid,draw=red},line width=0.6pt,forget plot]
  table[row sep=crcr]{%
2.5	0.00523865276527802\\
};
\end{axis}
\end{tikzpicture}%
		\caption{\gls{nmse}$_{\bx}$ for different $J$ in the second layer.}
		\label{figNMSExvsJ}
	\end{subfigure}
	\begin{subfigure}[htb]{0.47\linewidth}
		\centering
%
\definecolor{mycolor1}{rgb}{1.00000,0.00000,1.00000}%

\begin{tikzpicture}

\begin{axis}[%
width=\wfigtwo,
height=\hfigtwo,
at={(1.011in,0.642in)},
scale only axis,
unbounded coords=jump,
xmin=0.9,
xmax=4.1,
xtick={1.4,2.4,3.4},
xticklabels={{20},{50},{100}},
xlabel={J},
xmajorgrids,
ymin=0.01,
ymax=0.1,
ylabel={NMSE$_{\bz}$},
ymajorgrids,
yminorgrids,
ymode=log,
axis background/.style={fill=white}
]
\addplot [color=blue,dashed,line width=0.6pt,forget plot]
  table[row sep=crcr]{%
1.3	0.0350547216529075\\
1.3	0.0466816151165153\\
};
\addplot [color=blue,dashed,line width=0.6pt,forget plot]
  table[row sep=crcr]{%
2.3	0.0348351494225457\\
2.3	0.0394458695366301\\
};
\addplot [color=blue,dashed,line width=0.6pt,forget plot]
  table[row sep=crcr]{%
3.3	0.0328288143821155\\
3.3	0.0436338212183796\\
};
\addplot [color=blue,dashed,line width=0.6pt,forget plot]
  table[row sep=crcr]{%
1.3	0.0154338554198615\\
1.3	0.0263934273707967\\
};
\addplot [color=blue,dashed,line width=0.6pt,forget plot]
  table[row sep=crcr]{%
2.3	0.0200320592925306\\
2.3	0.0240711598633824\\
};
\addplot [color=blue,dashed,line width=0.6pt,forget plot]
  table[row sep=crcr]{%
3.3	0.0174410496999993\\
3.3	0.0210134632751294\\
};
\addplot [color=blue,solid,line width=0.6pt,forget plot]
  table[row sep=crcr]{%
1.255	0.0466816151165153\\
1.345	0.0466816151165153\\
};
\addplot [color=blue,solid,line width=0.6pt,forget plot]
  table[row sep=crcr]{%
2.255	0.0394458695366301\\
2.345	0.0394458695366301\\
};
\addplot [color=blue,solid,line width=0.6pt,forget plot]
  table[row sep=crcr]{%
3.255	0.0436338212183796\\
3.345	0.0436338212183796\\
};
\addplot [color=blue,solid,line width=0.6pt,forget plot]
  table[row sep=crcr]{%
1.255	0.0154338554198615\\
1.345	0.0154338554198615\\
};
\addplot [color=blue,solid,line width=0.6pt,forget plot]
  table[row sep=crcr]{%
2.255	0.0200320592925306\\
2.345	0.0200320592925306\\
};
\addplot [color=blue,solid,line width=0.6pt,forget plot]
  table[row sep=crcr]{%
3.255	0.0174410496999993\\
3.345	0.0174410496999993\\
};
\addplot [color=blue,solid,line width=0.6pt]
  table[row sep=crcr]{%
1.21	0.0263934273707967\\
1.21	0.0350547216529075\\
1.39	0.0350547216529075\\
1.39	0.0263934273707967\\
1.21	0.0263934273707967\\
};

\addplot [color=blue,solid,line width=0.6pt,forget plot]
  table[row sep=crcr]{%
2.21	0.0240711598633824\\
2.21	0.0348351494225457\\
2.39	0.0348351494225457\\
2.39	0.0240711598633824\\
2.21	0.0240711598633824\\
};
\addplot [color=blue,solid,line width=0.6pt,forget plot]
  table[row sep=crcr]{%
3.21	0.0210134632751294\\
3.21	0.0328288143821155\\
3.39	0.0328288143821155\\
3.39	0.0210134632751294\\
3.21	0.0210134632751294\\
};
\addplot [color=blue,solid,line width=0.6pt,forget plot]
  table[row sep=crcr]{%
1.21	0.030022757451455\\
1.39	0.030022757451455\\
};
\addplot [color=blue,solid,line width=0.6pt,forget plot]
  table[row sep=crcr]{%
2.21	0.0303779546081254\\
2.39	0.0303779546081254\\
};
\addplot [color=blue,solid,line width=0.6pt,forget plot]
  table[row sep=crcr]{%
3.21	0.0278919709331254\\
3.39	0.0278919709331254\\
};
\addplot [color=red,dashed,line width=0.6pt,forget plot]
  table[row sep=crcr]{%
1.5	0.0498049085328757\\
1.5	0.0699017276697274\\
};
\addplot [color=red,dashed,line width=0.6pt,forget plot]
  table[row sep=crcr]{%
2.5	0.0429635292764076\\
2.5	0.0521036571460623\\
};
\addplot [color=red,dashed,line width=0.6pt,forget plot]
  table[row sep=crcr]{%
3.5	0.0484490350585288\\
3.5	0.061090125033712\\
};
\addplot [color=red,dashed,line width=0.6pt,forget plot]
  table[row sep=crcr]{%
1.5	0.0211305865109231\\
1.5	0.0318585505755409\\
};
\addplot [color=red,dashed,line width=0.6pt,forget plot]
  table[row sep=crcr]{%
2.5	0.0253729461424452\\
2.5	0.0322288243294641\\
};
\addplot [color=red,dashed,line width=0.6pt,forget plot]
  table[row sep=crcr]{%
3.5	0.0266421109181879\\
3.5	0.0387006511334022\\
};
\addplot [color=red,solid,line width=0.6pt,forget plot]
  table[row sep=crcr]{%
1.455	0.0699017276697274\\
1.545	0.0699017276697274\\
};
\addplot [color=red,solid,line width=0.6pt,forget plot]
  table[row sep=crcr]{%
2.455	0.0521036571460623\\
2.545	0.0521036571460623\\
};
\addplot [color=red,solid,line width=0.6pt,forget plot]
  table[row sep=crcr]{%
3.455	0.061090125033712\\
3.545	0.061090125033712\\
};
\addplot [color=red,solid,line width=0.6pt,forget plot]
  table[row sep=crcr]{%
1.455	0.0211305865109231\\
1.545	0.0211305865109231\\
};
\addplot [color=red,solid,line width=0.6pt,forget plot]
  table[row sep=crcr]{%
2.455	0.0253729461424452\\
2.545	0.0253729461424452\\
};
\addplot [color=red,solid,line width=0.6pt,forget plot]
  table[row sep=crcr]{%
3.455	0.0266421109181879\\
3.545	0.0266421109181879\\
};
\addplot [color=red,solid,line width=0.6pt,forget plot]
  table[row sep=crcr]{%
1.41	0.0318585505755409\\
1.41	0.0498049085328757\\
1.59	0.0498049085328757\\
1.59	0.0318585505755409\\
1.41	0.0318585505755409\\
};
\addplot [color=red,solid,line width=0.6pt]
  table[row sep=crcr]{%
2.41	0.0322288243294641\\
2.41	0.0429635292764076\\
2.59	0.0429635292764076\\
2.59	0.0322288243294641\\
2.41	0.0322288243294641\\
};

\addplot [color=red,solid,line width=0.6pt,forget plot]
  table[row sep=crcr]{%
3.41	0.0387006511334022\\
3.41	0.0484490350585288\\
3.59	0.0484490350585288\\
3.59	0.0387006511334022\\
3.41	0.0387006511334022\\
};
\addplot [color=red,solid,line width=0.6pt,forget plot]
  table[row sep=crcr]{%
1.41	0.0399756573611275\\
1.59	0.0399756573611275\\
};
\addplot [color=red,solid,line width=0.6pt,forget plot]
  table[row sep=crcr]{%
2.41	0.0367597320910421\\
2.59	0.0367597320910421\\
};
\addplot [color=red,solid,line width=0.6pt,forget plot]
  table[row sep=crcr]{%
3.41	0.0437172283373171\\
3.59	0.0437172283373171\\
};
\addplot [color=black,only marks,mark=+,mark options={solid,draw=red},line width=0.6pt,forget plot]
  table[row sep=crcr]{%
2.5	0.0598383476050653\\
};
\addplot [color=black,only marks,mark=+,mark options={solid,draw=red},line width=0.6pt,forget plot]
  table[row sep=crcr]{%
3.5	0.0698525868697244\\
};

\end{axis}

\end{tikzpicture}%
		\caption{\gls{nmse}$_{\bz}$ for different $J$ in the second layer.}
		\label{figNMSEzvsJ}
	\end{subfigure}
 \hfill
	\begin{subfigure}[htb]{0.47\linewidth}
		\centering
%
%
\definecolor{mycolor1}{rgb}{0.00000,0.44700,0.74100}%
\definecolor{mycolor2}{rgb}{0.85000,0.32500,0.09800}%
\definecolor{mycolor3}{rgb}{1.00000,0.00000,1.00000}%
\begin{tikzpicture}

\begin{axis}[%
width=\wfigtwo,
height=\hfigtwo,
at={(1.011in,0.642in)},
scale only axis,
xmode=log,
xmin=20,
xmax=100,
xtick={20,50,100},
xticklabels={{20},{50},{100}},
ytick={0,2,4,6,8,10,12,14},
yticklabels={0,2,4,6,8,10,12,14},
xlabel={J},
xmajorgrids,
ymajorgrids,
xminorticks=true,
yminorticks=true,
ymin=0,
ymax=14,
ylabel={running time (hours)},
axis background/.style={fill=white},
legend style={legend cell align=left,align=left,draw=white!15!black}
]
\addplot [color=blue,solid,mark=diamond,line width=1pt,mark options={solid}]
  table[row sep=crcr]{%
20	2.71962832738889\\
50	6.42910987883333\\
100	12.7203321634444\\
};

\addplot [color=red,solid,mark=diamond,line width=1pt,mark options={solid}]
  table[row sep=crcr]{%
20	0.663131634888889\\
50	1.02566005238889\\
100	1.790963886\\
};


\end{axis}
\end{tikzpicture}%
		\caption{Running time for different $J$ in the second layer.}
		\label{figtimingvsJ}
	\end{subfigure}
	\caption{Averaged \gls{nmse}$_{\btheta}$ (\ref{figNMSEthetavsJ}), \gls{nmse}$_{\bx}$ (\ref{figNMSExvsJ}), \gls{nmse}$_{\bz}$ (\ref{figNMSEzvsJ}) and average running time (\ref{figtimingvsJ}) of \gls{smc}-\gls{smc}-\gls{ukf} (in blue) and \gls{smc}-\gls{enkf}-\gls{ekf} (in red), averaged over 50 simulation runs. The number of particles of the first layer (\gls{smc}) is set to $N=20$. We show results for different number of particles/ensembles $J$ in the second layer of the filter.}
	\label{fig:Multi changeJ}
\end{figure}

Figure \ref{fig:Multi changeN} compares the performance of the proposed methods and the \gls{enkf} for different values of $N$ (number of samples in the first layer) and $J=50$. This is shown with the averaged \gls{nmse}$_{\btheta}$, \gls{nmse}$_{\bx}$ and \gls{nmse}$_{\bz}$ together with the running time in hours. 
Similar to the previous figure, the first method (\gls{smc}-\gls{smc}-\gls{ukf}) shows a slight improvement in the accuracy of the slow state estimation as the number of samples $J$ increases (Fig. \ref{figNMSExvsN}). The second method (\gls{smc}-\gls{enkf}-\gls{ekf}) remains stable with $N$. The second method outperforms the first one in accuracy of the parameter estimation (Fig. \ref{figNMSEthetavsN}) and the slow state estimation (Fig. \ref{figNMSExvsN}), but not for the fast state estimation (Fig. \ref{figNMSEzvsN}). Again, the second method runs faster since the computational cost is considerably lower.

Finally, we show results for a computer experiment in which we have used the \gls{smc}-\gls{enkf}-\gls{ekf} method to estimate the parameters $F$, $C$, $B$ and $H$ and track the state variables of the two-scale Lorenz 96 system with dimension $d_x = 10$ and $d_z = 50$. The number of particles used to approximate the sequence of parameter posterior distributions is $N=50$ and the number of samples in the ensembles of the second layer is $J=50$.

Figure \ref{fig:Multi x1z1} shows the true state trajectories, together with their estimates, for the first slow state variable ($x_1$) and the first fast state variable ($z_1$) of the two-scale Lorenz 96 model. We note that although the accuracy of the estimation of the fast variable is similar throughout the whole simulation run (over 20 continuous-time units), we only show the last $2$ continuous-time units of the simulation.

In Fig. \ref{fig:Multi pdfs} we observe the estimated posterior \glspl{pdf} of the fixed parameters $F$, $C$, $B$ and $H$, together with the ground truth values. Figure \ref{subfig:Multi pdfF} displays the approximate posterior \gls{pdf} of the parameter $F$ (red dashed line) together with the true value $F=8$ (vertical black line), Fig. \ref{subfig:Multi pdfC} displays the approximate posterior \gls{pdf} of the parameter $C$ (blue dashed line) together with the true value $C=10$ (vertical black line), Fig. \ref{subfig:Multi pdfB} displays the approximate posterior \gls{pdf} of the parameter $B$ (green dashed line) together with the true value $B=15$ (vertical black line) and Fig. \ref{subfig:Multi pdfH} displays the approximate posterior \gls{pdf} of the parameter $H$ (magenta dashed line) together with the true value $H=0.75$ (vertical black line). 
We observe that for all the \glspl{pdf}, nearly all probability mass is allocated close to the true values, except for the parameter $B$ (Fig. \ref{subfig:Multi pdfB}). In this case, the \gls{pdf} is slightly shifted \gls{wrt} the true value.

\begin{figure}[t]
	\centering
	\begin{subfigure}[htb]{0.47\linewidth}
		\centering
		\input{figures2/NMSEthetavsN.tex}
		\caption{\gls{nmse}$_{\btheta}$ for different $N$ in the first layer.}
		\label{figNMSEthetavsN}
	\end{subfigure}
	\hfill
	\begin{subfigure}[htb]{0.47\linewidth}
		\centering
%
%
\definecolor{mycolor1}{rgb}{1.00000,0.00000,1.00000}%
\begin{tikzpicture}

\begin{axis}[%
width=\wfigtwo,
height=\hfigtwo,
at={(1.011in,0.642in)},
scale only axis,
unbounded coords=jump,
xmin=0.9,
xmax=4.1,
xtick={1.4,2.4,3.4},
xticklabels={{20},{50},{100}},
xlabel={N},
xmajorgrids,
ymin=0.0001,
ymax=0.02,
ylabel={NMSE$_{\bx}$},
ymajorgrids,
yminorgrids,
ymode=log,
axis background/.style={fill=white},
legend style={legend cell align=left,align=left,draw=white!15!black}
]
\addplot [color=blue,dashed,line width=0.6pt,forget plot]
  table[row sep=crcr]{%
1.3	0.00662836190083846\\
1.3	0.011349314243044\\
};
\addplot [color=blue,dashed,line width=0.6pt,forget plot]
  table[row sep=crcr]{%
2.3	0.00405176701343554\\
2.3	0.00630559852766459\\
};
\addplot [color=blue,dashed,line width=0.6pt,forget plot]
  table[row sep=crcr]{%
3.3	0.00368915893464494\\
3.3	0.00584254664651329\\
};
\addplot [color=blue,dashed,line width=0.6pt,forget plot]
  table[row sep=crcr]{%
1.3	0.00110311516414016\\
1.3	0.00245143294005704\\
};
\addplot [color=blue,dashed,line width=0.6pt,forget plot]
  table[row sep=crcr]{%
2.3	0.00150451591655049\\
2.3	0.00222503977442526\\
};
\addplot [color=blue,dashed,line width=0.6pt,forget plot]
  table[row sep=crcr]{%
3.3	0.00107924888075396\\
3.3	0.0018450693477211\\
};
\addplot [color=blue,solid,line width=0.6pt,forget plot]
  table[row sep=crcr]{%
1.255	0.011349314243044\\
1.345	0.011349314243044\\
};
\addplot [color=blue,solid,line width=0.6pt,forget plot]
  table[row sep=crcr]{%
2.255	0.00630559852766459\\
2.345	0.00630559852766459\\
};
\addplot [color=blue,solid,line width=0.6pt,forget plot]
  table[row sep=crcr]{%
3.255	0.00584254664651329\\
3.345	0.00584254664651329\\
};
\addplot [color=blue,solid,line width=0.6pt,forget plot]
  table[row sep=crcr]{%
1.255	0.00110311516414016\\
1.345	0.00110311516414016\\
};
\addplot [color=blue,solid,line width=0.6pt,forget plot]
  table[row sep=crcr]{%
2.255	0.00150451591655049\\
2.345	0.00150451591655049\\
};
\addplot [color=blue,solid,line width=0.6pt,forget plot]
  table[row sep=crcr]{%
3.255	0.00107924888075396\\
3.345	0.00107924888075396\\
};
\addplot [color=blue,line width=0.6pt,solid]
  table[row sep=crcr]{%
1.21	0.00245143294005704\\
1.21	0.00662836190083846\\
1.39	0.00662836190083846\\
1.39	0.00245143294005704\\
1.21	0.00245143294005704\\
};

\addplot [color=blue,solid,line width=0.6pt,forget plot]
  table[row sep=crcr]{%
2.21	0.00222503977442526\\
2.21	0.00405176701343554\\
2.39	0.00405176701343554\\
2.39	0.00222503977442526\\
2.21	0.00222503977442526\\
};
\addplot [color=blue,solid,line width=0.6pt,forget plot]
  table[row sep=crcr]{%
3.21	0.0018450693477211\\
3.21	0.00368915893464494\\
3.39	0.00368915893464494\\
3.39	0.0018450693477211\\
3.21	0.0018450693477211\\
};
\addplot [color=blue,solid,line width=0.6pt,forget plot]
  table[row sep=crcr]{%
1.21	0.00440737722958853\\
1.39	0.00440737722958853\\
};
\addplot [color=blue,solid,line width=0.6pt,forget plot]
  table[row sep=crcr]{%
2.21	0.00289775141942261\\
2.39	0.00289775141942261\\
};
\addplot [color=blue,solid,line width=0.6pt,forget plot]
  table[row sep=crcr]{%
3.21	0.00229155652281547\\
3.39	0.00229155652281547\\
};
\addplot [color=black,only marks,mark=+,mark options={solid,draw=red},line width=0.6pt,forget plot]
  table[row sep=crcr]{%
2.3	0.0220578810738787\\
2.3	0.055\\
};
\addplot [color=black,only marks,mark=+,mark options={solid,draw=red},line width=0.6pt,forget plot]
  table[row sep=crcr]{%
3.3	0.00796770248899538\\
};
\addplot [color=red,dashed,line width=0.6pt,forget plot]
  table[row sep=crcr]{%
1.5	0.00146485548468875\\
1.5	0.00239281805628899\\
};
\addplot [color=red,dashed,line width=0.6pt,forget plot]
  table[row sep=crcr]{%
2.5	0.00136707082084065\\
2.5	0.00179678891193301\\
};
\addplot [color=red,dashed,line width=0.6pt,forget plot]
  table[row sep=crcr]{%
3.5	0.00148254049007473\\
3.5	0.00193055230576486\\
};
\addplot [color=red,dashed,line width=0.6pt,forget plot]
  table[row sep=crcr]{%
1.5	0.000293204346510265\\
1.5	0.000597557886643933\\
};
\addplot [color=red,dashed,line width=0.6pt,forget plot]
  table[row sep=crcr]{%
2.5	0.000527116179840031\\
2.5	0.00093906202600462\\
};
\addplot [color=red,dashed,line width=0.6pt,forget plot]
  table[row sep=crcr]{%
3.5	0.000476159403632309\\
3.5	0.000652325377442363\\
};
\addplot [color=red,solid,line width=0.6pt,forget plot]
  table[row sep=crcr]{%
1.455	0.00239281805628899\\
1.545	0.00239281805628899\\
};
\addplot [color=red,solid,line width=0.6pt,forget plot]
  table[row sep=crcr]{%
2.455	0.00179678891193301\\
2.545	0.00179678891193301\\
};
\addplot [color=red,solid,line width=0.6pt,forget plot]
  table[row sep=crcr]{%
3.455	0.00193055230576486\\
3.545	0.00193055230576486\\
};
\addplot [color=red,solid,line width=0.6pt,forget plot]
  table[row sep=crcr]{%
1.455	0.000293204346510265\\
1.545	0.000293204346510265\\
};
\addplot [color=red,solid,line width=0.6pt,forget plot]
  table[row sep=crcr]{%
2.455	0.000527116179840031\\
2.545	0.000527116179840031\\
};
\addplot [color=red,solid,line width=0.6pt,forget plot]
  table[row sep=crcr]{%
3.455	0.000476159403632309\\
3.545	0.000476159403632309\\
};
\addplot [color=red,solid,line width=0.6pt,forget plot]
  table[row sep=crcr]{%
1.41	0.000597557886643933\\
1.41	0.00146485548468875\\
1.59	0.00146485548468875\\
1.59	0.000597557886643933\\
1.41	0.000597557886643933\\
};
\addplot [color=red,line width=0.6pt,solid]
  table[row sep=crcr]{%
2.41	0.00093906202600462\\
2.41	0.00136707082084065\\
2.59	0.00136707082084065\\
2.59	0.00093906202600462\\
2.41	0.00093906202600462\\
};

\addplot [color=red,solid,line width=0.6pt,forget plot]
  table[row sep=crcr]{%
3.41	0.000652325377442363\\
3.41	0.00148254049007473\\
3.59	0.00148254049007473\\
3.59	0.000652325377442363\\
3.41	0.000652325377442363\\
};
\addplot [color=red,solid,line width=0.6pt,forget plot]
  table[row sep=crcr]{%
1.41	0.00103802194880769\\
1.59	0.00103802194880769\\
};
\addplot [color=red,solid,line width=0.6pt,forget plot]
  table[row sep=crcr]{%
2.41	0.00114537246110696\\
2.59	0.00114537246110696\\
};
\addplot [color=red,solid,line width=0.6pt,forget plot]
  table[row sep=crcr]{%
3.41	0.000863099629542023\\
3.59	0.000863099629542023\\
};
\addplot [color=black,only marks,mark=+,mark options={solid,draw=red},line width=0.6pt,forget plot]
  table[row sep=crcr]{%
1.5	0.00523865276527802\\
};
\addplot [color=black,only marks,mark=+,mark options={solid,draw=red},line width=0.6pt,forget plot]
  table[row sep=crcr]{%
2.5	0.00265705455646395\\
2.5	0.00274985470810662\\
};
\addplot [color=black,only marks,mark=+,mark options={solid,draw=red},line width=0.6pt,forget plot]
  table[row sep=crcr]{%
3.5	0.00409951683665889\\
};
\end{axis}
\end{tikzpicture}%
		\caption{\gls{nmse}$_{\bx}$ for different $N$ in the first layer.}
		\label{figNMSExvsN}
	\end{subfigure}
	\begin{subfigure}[htb]{0.47\linewidth}
		\centering
%
%
\definecolor{mycolor1}{rgb}{1.00000,0.00000,1.00000}%
\begin{tikzpicture}

\begin{axis}[%
width=\wfigtwo,
height=\hfigtwo,
at={(1.011in,0.642in)},
scale only axis,
unbounded coords=jump,
xmin=0.9,
xmax=4.1,
xtick={1.4,2.4,3.4},
xticklabels={{20},{50},{100}},
xlabel={N},
xmajorgrids,
ymin=0.01,
ymax=0.1,
ylabel={NMSE$_{\bz}$},
ymajorgrids,
yminorgrids,
ymode=log,
axis background/.style={fill=white},
legend style={legend cell align=left,align=left,draw=white!15!black}
]
\addplot [color=blue,dashed,line width=0.6pt,forget plot]
  table[row sep=crcr]{%
1.3	0.0353536636417874\\
1.3	0.0423930324481931\\
};
\addplot [color=blue,dashed,line width=0.6pt,forget plot]
  table[row sep=crcr]{%
2.3	0.0301012266055526\\
2.3	0.0355157793861546\\
};
\addplot [color=blue,dashed,line width=0.6pt,forget plot]
  table[row sep=crcr]{%
3.3	0.0340188309773665\\
3.3	0.0411545533000865\\
};
\addplot [color=blue,dashed,line width=0.6pt,forget plot]
  table[row sep=crcr]{%
1.3	0.0200320592925306\\
1.3	0.0249383146057939\\
};
\addplot [color=blue,dashed,line width=0.6pt,forget plot]
  table[row sep=crcr]{%
2.3	0.0161586832914473\\
2.3	0.0218719810092368\\
};
\addplot [color=blue,dashed,line width=0.6pt,forget plot]
  table[row sep=crcr]{%
3.3	0.0192802097766585\\
3.3	0.0249029993117453\\
};
\addplot [color=blue,solid,line width=0.6pt,forget plot]
  table[row sep=crcr]{%
1.255	0.0423930324481931\\
1.345	0.0423930324481931\\
};
\addplot [color=blue,solid,line width=0.6pt,forget plot]
  table[row sep=crcr]{%
2.255	0.0355157793861546\\
2.345	0.0355157793861546\\
};
\addplot [color=blue,solid,line width=0.6pt,forget plot]
  table[row sep=crcr]{%
3.255	0.0411545533000865\\
3.345	0.0411545533000865\\
};
\addplot [color=blue,solid,line width=0.6pt,forget plot]
  table[row sep=crcr]{%
1.255	0.0200320592925306\\
1.345	0.0200320592925306\\
};
\addplot [color=blue,solid,line width=0.6pt,forget plot]
  table[row sep=crcr]{%
2.255	0.0161586832914473\\
2.345	0.0161586832914473\\
};
\addplot [color=blue,solid,line width=0.6pt,forget plot]
  table[row sep=crcr]{%
3.255	0.0192802097766585\\
3.345	0.0192802097766585\\
};
\addplot [color=blue,line width=0.6pt,solid]
  table[row sep=crcr]{%
1.21	0.0249383146057939\\
1.21	0.0353536636417874\\
1.39	0.0353536636417874\\
1.39	0.0249383146057939\\
1.21	0.0249383146057939\\
};

\addplot [color=blue,solid,line width=0.6pt,forget plot]
  table[row sep=crcr]{%
2.21	0.0218719810092368\\
2.21	0.0301012266055526\\
2.39	0.0301012266055526\\
2.39	0.0218719810092368\\
2.21	0.0218719810092368\\
};
\addplot [color=blue,solid,line width=0.6pt,forget plot]
  table[row sep=crcr]{%
3.21	0.0249029993117453\\
3.21	0.0340188309773665\\
3.39	0.0340188309773665\\
3.39	0.0249029993117453\\
3.21	0.0249029993117453\\
};
\addplot [color=blue,solid,line width=0.6pt,forget plot]
  table[row sep=crcr]{%
1.21	0.0313192682650768\\
1.39	0.0313192682650768\\
};
\addplot [color=blue,solid,line width=0.6pt,forget plot]
  table[row sep=crcr]{%
2.21	0.0253744160138051\\
2.39	0.0253744160138051\\
};
\addplot [color=blue,solid,line width=0.6pt,forget plot]
  table[row sep=crcr]{%
3.21	0.0293452412417603\\
3.39	0.0293452412417603\\
};

\addplot [color=red,dashed,line width=0.6pt,forget plot]
  table[row sep=crcr]{%
1.5	0.0429635292764076\\
1.5	0.0521036571460623\\
};
\addplot [color=red,dashed,line width=0.6pt,forget plot]
  table[row sep=crcr]{%
2.5	0.0500403606394873\\
2.5	0.0677087800991528\\
};
\addplot [color=red,dashed,line width=0.6pt,forget plot]
  table[row sep=crcr]{%
3.5	0.0494588461203734\\
3.5	0.0575186597664304\\
};
\addplot [color=red,dashed,line width=0.6pt,forget plot]
  table[row sep=crcr]{%
1.5	0.0253729461424452\\
1.5	0.0322288243294641\\
};
\addplot [color=red,dashed,line width=0.6pt,forget plot]
  table[row sep=crcr]{%
2.5	0.0231162178635405\\
2.5	0.0344310368833181\\
};
\addplot [color=red,dashed,line width=0.6pt,forget plot]
  table[row sep=crcr]{%
3.5	0.034778574156388\\
3.5	0.0409111808086439\\
};
\addplot [color=red,solid,line width=0.6pt,forget plot]
  table[row sep=crcr]{%
1.455	0.0521036571460623\\
1.545	0.0521036571460623\\
};
\addplot [color=red,solid,line width=0.6pt,forget plot]
  table[row sep=crcr]{%
2.455	0.0677087800991528\\
2.545	0.0677087800991528\\
};
\addplot [color=red,solid,line width=0.6pt,forget plot]
  table[row sep=crcr]{%
3.455	0.0575186597664304\\
3.545	0.0575186597664304\\
};
\addplot [color=red,solid,line width=0.6pt,forget plot]
  table[row sep=crcr]{%
1.455	0.0253729461424452\\
1.545	0.0253729461424452\\
};
\addplot [color=red,solid,line width=0.6pt,forget plot]
  table[row sep=crcr]{%
2.455	0.0231162178635405\\
2.545	0.0231162178635405\\
};
\addplot [color=red,solid,line width=0.6pt,forget plot]
  table[row sep=crcr]{%
3.455	0.034778574156388\\
3.545	0.034778574156388\\
};
\addplot [color=red,solid,line width=0.6pt,forget plot]
  table[row sep=crcr]{%
1.41	0.0322288243294641\\
1.41	0.0429635292764076\\
1.59	0.0429635292764076\\
1.59	0.0322288243294641\\
1.41	0.0322288243294641\\
};
\addplot [color=red,line width=0.6pt,solid]
  table[row sep=crcr]{%
2.41	0.0344310368833181\\
2.41	0.0500403606394873\\
2.59	0.0500403606394873\\
2.59	0.0344310368833181\\
2.41	0.0344310368833181\\
};

\addplot [color=red,solid,line width=0.6pt,forget plot]
  table[row sep=crcr]{%
3.41	0.0409111808086439\\
3.41	0.0494588461203734\\
3.59	0.0494588461203734\\
3.59	0.0409111808086439\\
3.41	0.0409111808086439\\
};
\addplot [color=red,solid,line width=0.6pt,forget plot]
  table[row sep=crcr]{%
1.41	0.0367597320910421\\
1.59	0.0367597320910421\\
};
\addplot [color=red,solid,line width=0.6pt,forget plot]
  table[row sep=crcr]{%
2.41	0.0427704721061045\\
2.59	0.0427704721061045\\
};
\addplot [color=red,solid,line width=0.6pt,forget plot]
  table[row sep=crcr]{%
3.41	0.043236529048895\\
3.59	0.043236529048895\\
};
\addplot [color=black,only marks,mark=+,mark options={solid,draw=red},line width=0.6pt,forget plot]
  table[row sep=crcr]{%
1.5	0.0598383476050653\\
};

\addplot [color=black,only marks,mark=+,mark options={solid,draw=red},line width=0.6pt,forget plot]
  table[row sep=crcr]{%
3.5	0.027620341960217\\
};

\end{axis}
\end{tikzpicture}%
		\caption{\gls{nmse}$_{\bz}$ for different $N$ in the first layer.}
		\label{figNMSEzvsN}
	\end{subfigure}
	\hfill
	\begin{subfigure}[htb]{0.47\linewidth}
		\centering
%
%
\definecolor{mycolor1}{rgb}{1.00000,0.00000,1.00000}%
\begin{tikzpicture}

\begin{axis}[%
width=\wfigtwo,
height=\hfigtwo,
at={(1.011in,0.642in)},
scale only axis,
xtick={20,50,100},
xticklabels={{20},{50},{100}},
ytick={0,5,10,15,20,25,30,35},
yticklabels={0,,10,,20,,30,},
xmajorgrids,
ymajorgrids,
yminorticks=true,
ymin=0,
xmode=log,
xmin=20,
xmax=100,
xminorticks=true,
xlabel={N},
ymax=35,
ylabel={running time (hours)},
axis background/.style={fill=white},
legend style={legend cell align=left,align=left,draw=white!15!black}
]
\addplot [color=blue,solid,mark=diamond,line width=1pt,mark options={solid}]
  table[row sep=crcr]{%
20	6.50940621519444\\
50	16.1772234553611\\
100	32.7815791433056\\
};

\addplot [color=red,solid,mark=diamond,line width=1pt,mark options={solid}]
  table[row sep=crcr]{%
20	1.02275264002778\\
50	2.35986439433333\\
100	5.26847848034722\\
};

\end{axis}
\end{tikzpicture}%
		\caption{Running time for different $N$ in the first layer.}
		\label{figtimingvsN}
	\end{subfigure}
	\caption{Averaged \gls{nmse}$_{\btheta}$ (\ref{figNMSEthetavsJ}), \gls{nmse}$_{\bx}$ (\ref{figNMSExvsJ}), \gls{nmse}$_{\bz}$ (\ref{figNMSEzvsJ}) and average running time (\ref{figtimingvsJ}) of \gls{smc}-\gls{smc}-\gls{ukf} (in blue) and \gls{smc}-\gls{enkf}-\gls{ekf} (in red), averaged over 50 simulation runs. The number of particles/ensembles of the second layer (\gls{smc} for the first method and \gls{enkf} for the second method) is set to $J=50$. We show results for different number of particles $N$ in the first layer of the filter.}
	\label{fig:Multi changeN}
\end{figure}

\begin{figure}[h!]
\centering
	\begin{subfigure}[htb]{0.47\linewidth}
	\centering
	    \input{figures2/x1.tex}
		\caption{Time sequence of $x_1$}
		\label{subfig:Multi x1}
	\end{subfigure}
	\begin{subfigure}[htb]{0.47\linewidth}
	\centering
	    \input{figures2/z1.tex}
		\caption{Time sequence of $x_2$}
		\label{subfig:Multi z1}
	\end{subfigure}
	\caption{Sequences of state values (black line) and estimates (dashed red line) in $x_1$ (plot \ref{subfig:Multi x1}) and $z_1$ (plot \ref{subfig:Multi z1}) over time.}
	\label{fig:Multi x1z1}
\end{figure}

\begin{figure}[h!]
\vspace{1cm}
\centering
	\begin{subfigure}[htb]{0.47\linewidth}
	\centering
%
%
\begin{tikzpicture}

\begin{axis}[%
width=\wfigtwo,
height=\hfigtwo,
at={(1.011in,0.642in)},
scale only axis,
xmin=7,
xmax=9.5,
xlabel={$F$},
ymin=0,
ymax=1.15355206287617,
ylabel={posterior density},
axis background/.style={fill=white},
legend style={legend cell align=left,align=left,draw=white!15!black}
]
\addplot [color=red,dashed,line width=2.0pt,forget plot]
  table[row sep=crcr]{%
7	0.0103223607324786\\
7.05	0.0152845703352716\\
7.1	0.0222421361241479\\
7.15	0.0318088489427385\\
7.2	0.0447060951237465\\
7.25	0.0617492880969335\\
7.3	0.0838190429373302\\
7.35	0.111814478765395\\
7.4	0.146587540140091\\
7.45	0.188859480012903\\
7.5	0.239123512368332\\
7.55	0.297540823981874\\
7.6	0.363840173397465\\
7.65	0.437233629906206\\
7.7	0.516362019188029\\
7.75	0.599282841286937\\
7.8	0.683510527494907\\
7.85	0.766113952341846\\
7.9	0.843869551668219\\
7.95	0.913461028990749\\
8	0.971709546638355\\
8.05	1.01581268210345\\
8.1	1.04356734939338\\
8.15	1.05355206287617\\
8.2	1.04524756611925\\
8.25	1.0190815778644\\
8.3	0.976392288034816\\
8.35	0.91931494035742\\
8.4	0.850604883295444\\
8.45	0.773417497555913\\
8.5	0.691069426558536\\
8.55	0.60680609487206\\
8.6	0.523597741930736\\
8.65	0.443980797483369\\
8.7	0.369954421627861\\
8.75	0.302934614550965\\
8.8	0.243761581360385\\
8.85	0.19275086797164\\
8.9	0.149775647507362\\
8.95	0.114366521454513\\
9	0.0858160524424522\\
9.05	0.0632774764638211\\
9.1	0.0458500546262983\\
9.15	0.0326467350125394\\
9.2	0.022842727198151\\
9.25	0.0157059270896182\\
9.3	0.0106117194308445\\
9.35	0.00704552958437339\\
9.4	0.00459670236138768\\
9.45	0.00294702044049525\\
9.5	0.00185662037781566\\
9.55	0.00114938484540528\\
9.6	0.000699212863605889\\
9.65	0.000417978701778858\\
9.7	0.000245526808740298\\
9.75	0.000141723850356007\\
9.8	8.03869348601568e-05\\
9.85	4.48049251535427e-05\\
9.9	2.45393509408538e-05\\
9.95	1.32067724625278e-05\\
10	6.9843487876225e-06\\
};
\addplot [color=black,solid,line width=1.0pt,mark=o,mark options={solid},forget plot]
  table[row sep=crcr]{%
8	0\\
8	1.05355206287617\\
};
\end{axis}
\end{tikzpicture}%
		\caption{Posterior \gls{pdf} of $F$}
		\label{subfig:Multi pdfF}
	\end{subfigure}
	\hfill
	\begin{subfigure}[htb]{0.47\linewidth}
	\centering
%
%
\begin{tikzpicture}

\begin{axis}[%
width=\wfigtwo,
height=\hfigtwo,
at={(1.011in,0.642in)},
scale only axis,
xmin=8.5,
xmax=11.5,
xlabel={$C$},
ymin=0,
ymax=1.14443833799968,
ylabel={posterior density},
xtick={8.5,9,9.5,10,10.5,11,11.5},
xticklabels={,9,,10,,11,},
axis background/.style={fill=white},
legend style={legend cell align=left,align=left,draw=white!15!black}
]
\addplot [color=blue,dashed,line width=2.0pt,forget plot]
  table[row sep=crcr]{%
8	1.20866970022535e-05\\
8.05	2.2452775571382e-05\\
8.1	4.09891512180586e-05\\
8.15	7.35370911282112e-05\\
8.2	0.000129653665436219\\
8.25	0.000224649906055316\\
8.3	0.00038253668639636\\
8.35	0.000640158814716269\\
8.4	0.00105281629551852\\
8.45	0.00170164985946522\\
8.5	0.00270298173395583\\
8.55	0.00421962598894624\\
8.6	0.00647389192056538\\
8.65	0.00976158487932866\\
8.7	0.0144657668346568\\
8.75	0.0210684079853113\\
8.8	0.0301574124891397\\
8.85	0.0424259484528996\\
8.9	0.058660703379219\\
8.95	0.0797157900420758\\
9	0.106469704529509\\
9.05	0.139764100688396\\
9.1	0.180325214553768\\
9.15	0.228671438303892\\
9.2	0.28501354060696\\
9.25	0.349156942876923\\
9.3	0.420417762461177\\
9.35	0.497565465051476\\
9.4	0.578804444987809\\
9.45	0.661804381494005\\
9.5	0.743784805042836\\
9.55	0.821653311332377\\
9.6	0.8921899923101\\
9.65	0.952263894784932\\
9.7	0.999061763989244\\
9.75	1.03030599615969\\
9.8	1.04443833799968\\
9.85	1.04074870704958\\
9.9	1.01943430413011\\
9.95	0.981582169097755\\
10	0.929077325445439\\
10.05	0.864447316588905\\
10.1	0.790660970048056\\
10.15	0.710903644841174\\
10.2	0.628352497272967\\
10.25	0.545973452710968\\
10.3	0.466357108372366\\
10.35	0.391604598397021\\
10.4	0.323267594212579\\
10.45	0.262340136247648\\
10.5	0.209294746535525\\
10.55	0.164151791692692\\
10.6	0.126569533113431\\
10.65	0.0959425712077752\\
10.7	0.0714980755085531\\
10.75	0.0523817727725826\\
10.8	0.0377286008312388\\
10.85	0.0267157580623994\\
10.9	0.0185982474154084\\
10.95	0.0127287415028335\\
11	0.00856463446446382\\
11.05	0.00566556056896799\\
11.1	0.00368458156131724\\
11.15	0.00235583575553444\\
11.2	0.00148085819333216\\
11.25	0.000915152381391753\\
11.3	0.00055601236403476\\
11.35	0.00033211306415202\\
11.4	0.000195028057895555\\
11.45	0.000112594397094013\\
11.5	6.39063046956408e-05\\
11.55	3.56595901202552e-05\\
11.6	1.95619867084481e-05\\
11.65	1.0549978821671e-05\\
11.7	5.59358280657438e-06\\
11.75	2.91558732207039e-06\\
11.8	1.49402197061874e-06\\
11.85	7.52627590950208e-07\\
11.9	3.72728502538258e-07\\
11.95	1.8146435373953e-07\\
12	8.6850651756428e-08\\
};
\addplot [color=black,solid,line width=1.0pt,mark=o,mark options={solid},forget plot]
  table[row sep=crcr]{%
10	0\\
10	1.04443833799968\\
};
\end{axis}
\end{tikzpicture}%
		\caption{Posterior pdf of $C$}
		\label{subfig:Multi pdfC}
	\end{subfigure}
	\begin{subfigure}[htb]{0.47\linewidth}
	\centering
%
%
\begin{tikzpicture}

\begin{axis}[%
width=\wfigtwo,
height=\hfigtwo,
at={(1.011in,0.642in)},
scale only axis,
xmin=13,
xmax=16,
xlabel={$B$},
ymin=0,
ymax=1.12482413575083,
ylabel={posterior density},
xtick={13,13.5,14,14.5,15,15.5,16},
xticklabels={13,,14,,15,,16},
axis background/.style={fill=white},
legend style={legend cell align=left,align=left,draw=white!15!black}
]
\addplot [color=green,dashed,line width=2.0pt,forget plot]
  table[row sep=crcr]{%
12	3.01527634531643e-08\\
12.05	6.44387634531308e-08\\
12.1	1.35348717267035e-07\\
12.15	2.79418095549703e-07\\
12.2	5.66963646103973e-07\\
12.25	1.13074157082771e-06\\
12.3	2.21659542088468e-06\\
12.35	4.27102586522974e-06\\
12.4	8.08925910198994e-06\\
12.45	1.50599819397159e-05\\
12.5	2.75605579992375e-05\\
12.55	4.95802146987323e-05\\
12.6	8.76788667752617e-05\\
12.65	0.000152425333374035\\
12.7	0.000260498320038755\\
12.75	0.000437671589041798\\
12.8	0.000722932570174781\\
12.85	0.00117398840470568\\
12.9	0.00187437799337472\\
12.95	0.00294231281302973\\
13	0.00454119228066047\\
13.05	0.00689146429843019\\
13.1	0.0102831213181397\\
13.15	0.0150876472190654\\
13.2	0.0217676956904225\\
13.25	0.0308822521022664\\
13.3	0.043084604873572\\
13.35	0.0591102516251443\\
13.4	0.0797520238026156\\
13.45	0.105820352854191\\
13.5	0.138087800941679\\
13.55	0.177218743718144\\
13.6	0.223687322308902\\
13.65	0.277689257376242\\
13.7	0.339055507344421\\
13.75	0.407177641013685\\
13.8	0.480955744463406\\
13.85	0.558779312605694\\
13.9	0.638549650528188\\
13.95	0.717748814105972\\
14	0.793555307974333\\
14.05	0.863001157398559\\
14.1	0.923159321578108\\
14.15	0.971345573314996\\
14.2	1.00531575571106\\
14.25	1.02343837685023\\
14.3	1.02482413575083\\
14.35	1.00939810405049\\
14.4	0.977906424412893\\
14.45	0.931856693817029\\
14.5	0.87339863887223\\
14.55	0.805158196923007\\
14.6	0.730042785960251\\
14.65	0.651037785143895\\
14.7	0.571013865607697\\
14.75	0.492562032895513\\
14.8	0.417868643395503\\
14.85	0.348637039943658\\
14.9	0.286056691285939\\
14.95	0.230815611573098\\
15	0.183147974478824\\
15.05	0.142906546897913\\
15.1	0.109648892382403\\
15.15	0.0827270344092059\\
15.2	0.0613720559885459\\
15.25	0.0447674988938351\\
15.3	0.0321079749476574\\
15.35	0.0226417502201534\\
15.4	0.015697961684371\\
15.45	0.0107004507952713\\
15.5	0.00717093782373389\\
15.55	0.00472448695035111\\
15.6	0.00306004726830529\\
15.65	0.00194843812213953\\
15.7	0.00121960889279807\\
15.75	0.000750448304606942\\
15.8	0.000453918882974748\\
15.85	0.000269888460106809\\
15.9	0.000157735918309071\\
15.95	9.06168615828602e-05\\
16	5.11695721289608e-05\\
16.05	2.84008551282584e-05\\
16.1	1.54938988316179e-05\\
16.15	8.30792906944098e-06\\
16.2	4.37845601133028e-06\\
16.25	2.26797890266436e-06\\
16.3	1.15462451295924e-06\\
16.35	5.77724095664685e-07\\
16.4	2.84100889778409e-07\\
16.45	1.37306728819718e-07\\
16.5	6.52188465271857e-08\\
16.55	3.0444697086969e-08\\
16.6	1.39669912657703e-08\\
16.65	6.29712553528224e-09\\
16.7	2.79013976101484e-09\\
16.75	1.21492568063727e-09\\
16.8	5.19888047064216e-10\\
16.85	2.1862645833047e-10\\
16.9	9.03493318851929e-11\\
16.95	3.6692157357636e-11\\
17	1.46434874263752e-11\\
};
\addplot [color=black,solid,line width=1.0pt,mark=o,mark options={solid},forget plot]
  table[row sep=crcr]{%
15	0\\
15	1.02482413575083\\
};
\end{axis}
\end{tikzpicture}%
		\caption{Posterior \gls{pdf} of $B$}
		\label{subfig:Multi pdfB}
	\end{subfigure}
	\hfill
	\begin{subfigure}[htb]{0.47\linewidth}
	\centering
%
%
\definecolor{mycolor1}{rgb}{1.00000,0.00000,1.00000}%
\begin{tikzpicture}

\begin{axis}[%
width=\wfigtwo,
height=\hfigtwo,
at={(1.011in,0.642in)},
scale only axis,
xmin=-0.5,
xmax=2.5,
xlabel={$H$},
ymin=0,
ymax=1.11069307401767,
ylabel={posterior density},
xtick={-0.5,0,0.5,1,1.5,2,2.5},
xticklabels={,0,,1,,2,},
axis background/.style={fill=white},
legend style={legend cell align=left,align=left,draw=white!15!black}
]
\addplot [color=mycolor1,dashed,line width=2.0pt,forget plot]
  table[row sep=crcr]{%
-0.5	0.00108791282699081\\
-0.45	0.00173701719439293\\
-0.4	0.00272670012534656\\
-0.35	0.0042083395392705\\
-0.3	0.00638619075563091\\
-0.25	0.00952903614733021\\
-0.2	0.0139814083835426\\
-0.15	0.0201728472448714\\
-0.1	0.0286231754784925\\
-0.05	0.0399413945605474\\
0	0.0548156131603486\\
0.05	0.0739915452654493\\
0.1	0.0982376561247802\\
0.15	0.128296060672342\\
0.2	0.164819795386597\\
0.25	0.208299008185783\\
0.3	0.258980760405416\\
0.35	0.31678923445969\\
0.4	0.381254846725496\\
0.45	0.451461712939402\\
0.5	0.526022780792937\\
0.55	0.603090520919014\\
0.6	0.680408315818411\\
0.65	0.755403781866249\\
0.7	0.825320595250579\\
0.75	0.887380542608416\\
0.8	0.938963160503632\\
0.85	0.977787143771803\\
0.9	1.00207625297453\\
0.95	1.01069307401767\\
1	1.00322671918779\\
1.05	0.980025127138096\\
1.1	0.94216844850212\\
1.15	0.891386318078864\\
1.2	0.829927757436133\\
1.25	0.760397225553079\\
1.3	0.685573329525618\\
1.35	0.608227593713651\\
1.4	0.530959459289251\\
1.45	0.45606065068851\\
1.5	0.385417745341965\\
1.55	0.320456898661305\\
1.6	0.26212990459477\\
1.65	0.210936715661517\\
1.7	0.166976630150381\\
1.75	0.130018782323511\\
1.8	0.0995823273703399\\
1.85	0.0750175960646333\\
1.9	0.0555811822115872\\
1.95	0.0405000462722373\\
2	0.0290219144968229\\
2.05	0.0204512348988587\\
2.1	0.0141715257383323\\
2.15	0.00965602771996106\\
2.2	0.00646914767330076\\
2.25	0.00426132634887829\\
2.3	0.00275978264583234\\
2.35	0.00175720002769543\\
2.4	0.00109993838079849\\
2.45	0.00067686436811376\\
2.5	0.000409455875606326\\
};
\addplot [color=black,solid,line width=1.0pt,mark=o,mark options={solid},forget plot]
  table[row sep=crcr]{%
0.75	0\\
0.75	1.01069307401767\\
};
\end{axis}
\end{tikzpicture}%
		\caption{Posterior pdf of $H$}
		\label{subfig:Multi pdfH}
	\end{subfigure}
	\caption{Posterior density of the parameters (dashed lines) at time $\tau = 20$. The true values are indicated by a black vertical line.} 
	\label{fig:Multi pdfs}
\end{figure}

\section{Conclusions} \label{sec:Multi Conclusions}

We have introduced a further generalization of the \gls{nhf} methodology of \cite{Perez-Vieites18} that, using long sequences of observations collected over time, estimates the static parameters and the stochastic dynamical variables of a class of heterogeneous multi-scale state-space models \cite{Abdulle2012}.
This scheme combines three layers of filters, one inside the other. It approximates recursively the posterior probability distributions of the parameters and the two sets of state variables given the sequence of available observations. 
In a first layer of computation we approximate the posterior probability distribution of the parameters, in a second layer we approximate the posterior probability distribution of the slow state variables, and the posterior probability distribution of the fast state variables is approximated in a third layer. The inference techniques used in each layer can vary, leading to different computational costs and degrees of accuracy.
To be specific, we describe two possible algorithms that derive from this scheme, combining Monte Carlo methods and Gaussian filters at different layers. The first method involves using sequential Monte Carlo (SMC) methods in both first and second layers, together with a bank of unscented Kalman filter (UKFs) in the third layer (i.e., the \gls{smc}-\gls{smc}-\gls{ukf}). The second method employs a \gls{smc} in the first layer, ensemble Kalman filters (EnKFs) at the second layer and introduces the use of a bank of extended Kalman filters (EKFs) in the third layer (i.e., the \gls{smc}-\gls{enkf}-\gls{ekf}). 
We have presented numerical results for a two-scale stochastic Lorenz 96 model with synthetic data and we have evaluated the performance of the algorithm in terms of the normalized mean square errors (NMSEs) for the parameters and the dynamic (slow and fast) state variables.
The proposed implementations (both of them) obtain good results in terms of accuracy, having a considerably reduction in running time (i.e., the computational cost) with the second method. Further research is still needed, studying the stability of the multi-layer structure when the sequence of observations are rare and/or few.
Moreover, we can compare the proposed algorithms with other methods, using the Lorenz 96 system but also other models.

%
\section*{Acknowledgements}

This work was partially supported by the Research Council of Finland Flagship programme: Finnish Center for Artificial Intelligence FCAI. JM acknowledges the support of the Office of Naval Research (award N00014-22-1-2647) and Spain's \textit{Agencia Estatal de Investigaci\'on} (ref. PID2021-125159NB-I00 TYCHE) funded by MCIN/AEI/10.13039/501100011033 and by ``ERDF A way of making Europe".

\appendix
\section{Implementation of SMC-SMC-UKF}\label{ap:1}

Algorithm \ref{alg:Multi FirstScheme SMC2} describes the implementation of a bank of conditional \gls{smc} schemes in the second layer of the multi-scale nested smoother, one for each particle $\widetilde{\btheta}_{t}^{(i)}$, $i=1, \ldots, N$. In this second layer we approximate the posterior distribution with density $p(\bx_{0:t} | \by_{1:t}, \widetilde{\btheta}_{t}^{(i)})$. The procedure is similar to the one in Algorithm \ref{alg:Multi FirstScheme SMC1}.
It receives as input the prior \gls{pdf} of the slow state variables, $p(\bx_0)$. In the initialization step, the prior \gls{pdf} is used to generate $J$ \gls{iid} Monte Carlo slow state particles per each parameter sample, $\{\bx_0^{(i,j)}\}_{j=1}^J$. This implies that the whole procedure is conditioned on a fixed value of the index $i$ and on a specific vector of parameters $\widetilde{\btheta}_{t}^{(i)}$.

At every time step $t$, we run $J$ filters in the third layer to generate $J$ particles $\widetilde{\bx}_{t}^{(i,j)}$. Within the third layer of computation we also compute the approximate likelihood 
$$
\widehat{p}(\by_{t} | \widetilde{\bx}_{t}^{(i,j)}, \bx_{0:t-1}^{(i,j)}, \by_{1:t-1}, \widetilde{\btheta}_{t}^{(i)}) \approx
p(\by_{t} | \widetilde{\bx}_{t}^{(i,j)}, \bx_{0:t-1}^{(i,j)}, \by_{1:t-1}, \widetilde{\btheta}_{t}^{(i)})
$$ 
and the non-normalized weights $\{\widetilde{u}_{t}^{(i,j)}\}_{j=1}^J$ in step \ref{stepal2likelihood}. Additionally, the fast state variables are propagated in the third layer of computation (as described later in this section), obtaining a new set of points $\{ {\bz}_{ht|ht}^{(i,j,l)}\}_{l=0}^{L-1}$ at time $t$.

\newpage
\par\noindent\rule{0.9\textwidth}{0.4pt}
\begin{algoritmo} {\gls{smc} approximation of $p(\bx_{0:t} | \by_{1:t}, \btheta)$ \label{alg:Multi FirstScheme SMC2}} 
		
		\textbf{Inputs}
		\begin{itemize}
            \item[-] Prior distribution $p(\bx_0)$.
			\item[-] Known parameter $\widetilde{\btheta}_{t}^{(i)}$, i.e., one fixed value for index $i$. 
   
            \item[-] If $t>1$, known initial states $\{ \bx_{t-1}^{(i,j)}, \{ \bz_{h(t-1)|h(t-1)}^{(i,j,l)} \}_{l=0}^{L-1} \}_{j=1}^J $.

		\end{itemize}

        \textbf{Initialization}
        \begin{itemize}
			         \item[-] If $t=1$, draw $J$ i.i.d. samples $\bx_0^{(i,j)} \sim p(\bx_0)$, for $j = 1,\ldots,J$.
        \end{itemize}

		\textbf{Procedure} For any $t> 0$:
		\begin{enumerate}
			\item For $j=1,\ldots,J$, run layer 3 algorithm to:
			\begin{enumerate} 
				\item Generate a new particle $\widetilde{\bx}_{t}^{(i,j)}$.
				\item Compute the approximate likelihood $\widetilde{u}_{t}^{(i,j)} = \widehat{p}^{L} (\by_{t} | \widetilde{\bx}_{t}^{(i,j)}, \bx_{0:t-1}^{(i,j)}, \by_{1:t-1}, \widetilde{\btheta}_{t}^{(i)})$. \label{stepal2likelihood}
				\item Obtain new particles $\{ {\bz}_{ht|ht}^{(i,j,l)}\}_{l=0}^{L-1}$ at time $t$.
                \label{stepal2propagatestate}

			\end{enumerate}
            \item Compute the unnormalized weights for layer 1 (Alg. \ref{alg:Multi FirstScheme SMC1}) \label{stepal2weigthlayer1}
            \begin{equation}
                \widetilde{v}_{t}^{(i)} = \frac{1}{J} \sum_{j=1}^{J} \widetilde{u}_{t}^{(i,j)}.
            \end{equation}

            \item Normalize the weights \label{stepal2normweights}
				\begin{equation}
				u_{t}^{(i,j)} = \frac{\widetilde{u}_{t}^{(i,j)}}{\sum_{k=1}^{J} \widetilde{u}_{t}^{(i,k)}}. \label{eqal2normweights}
				\end{equation}
   
			\item Resample: draw indices $m_1, \ldots, m_J$ with probability $u_{t}^{(i,1)}, \ldots, u_{t}^{(i,J)}$, and for $j=1,\ldots,J$ set \label{stepal2resampling}
			\begin{equation}
			\{\bx_{t}^{(i,j)}, \{{\bz}_{ht|ht}^{(i,j,l)}\}_{l=0}^{L-1} \} = \{\widetilde{\bx}_{t}^{(i,m_j)}, \{{\bz}_{ht|ht}^{(i,m_j,l)}\}_{l=0}^{L-1} \}.
			\end{equation}

		\end{enumerate}
		
		\textbf{Outputs:} $\{\bx_{t}^{(i,j)}, \{{\bz}_{ht|ht}^{(i,j,l)}\}_{l=0}^{L-1} \}_{j=1}^J $ and $\widetilde{v}_{t}^{(i)}$.
		
\end{algoritmo}
\par\noindent\rule{0.9\textwidth}{0.4pt}
\vspace{2mm}

With the outputs provided by the third layer (namely, the set of particles and points $\{ \widetilde{\bx}_{t}^{(i,j)}, \{ \bz_{ht|ht}^{(i,j,l)} \}_{l=0}^{L-1} \}_{j=1}^J$, and the weights $\{\widetilde{u}_{t}^{(i,j)}\}_{j=1}^J$) we can obtain the weights $\widetilde{v}_{t}^{(i)}$. These weights, that are needed in the first layer, are computed by averaging the weights $\{\widetilde{u}_{t}^{(i,j)}\}_{j=1}^J$ in step \ref{stepal2weigthlayer1}.\footnote{The average $\widetilde{v}_{t}^{(i)} = \frac{1}{J}\sum_{j=1}^J\widetilde{u}_{t}^{(i,j)}$ is an approximation of the integral of Eq. \eqref{eq:Multi likelihood 1st}.} 
After, one can use the normalized weights $\{ u_{t}^{(i,j)} \}_{j=1}^J$ obtained in step \ref{stepal2normweights}
to resample the set $\{ \bx_{t}^{(i,j)}, \{ \bz_{ht|ht}^{(i,j,l)} \}_{l=0}^{L-1} \}_{j=1}^J$.
Note that these sets of samples or points describe the approximate distributions $\widehat{p}(\bx_{0:t}| \by_{1:t}, \widetilde{\btheta}_t^{(i)})$ and $\widehat{p}(\bz_{h(t-1)+1:ht} | \bx_{0:t}^{(i,j)}, \by_{1:t}, \widetilde{\btheta}_t^{(i)})$, respectively.

Algorithm \ref{alg:Multi FirstScheme UKF} outlines the implementation of a bank of \glspl{ukf} \cite{Julier04} conditional on each parameter sample $\widetilde{\btheta}_{t}^{(i)}$ and the set of slow states $\{ \widetilde{\bx}_{t}^{(i,j)} \} \cup \bx_{0:t-1}^{(i,j)}$. Thus, the indices $i$ and $j$ are considered fixed in the description of the algorithm. If we follow the template of the optimal smoother, then we should seek an approximation of the density $p(\bz_{{h(t-1)+1:ht}} | \bx_{0:t-1}^{(i,j)},\by_{1:t-1}, \widetilde{\btheta}_{t}^{(i)})$. However, performing this calculation with a \gls{ukf}-like scheme implies that the dimension of the filter should be $d_z \times h$, in order to include the whole subsequence of states $\bz_{{h(t-1)+1:ht}}$. Such approach would demand $2 d_z h + 1$ sigma-points for each conditional \gls{ukf} algorithm, and the computation of $NJL$ covariance matrices with dimension $2d_zh \times 2d_zh$ each, which is impractical even for moderate $d_z$ and $h$. For simplicity, in order to avoid operations with large matrices, we choose to compute Gaussian approximations of the marginal predictive densities 
$
p(\bz_q|\bx_{0:t-1}^{(i,j)},\by_{1:t-1},\widetilde{\btheta}_{t}^{(i)}), 
$
for $q = h(t-1)+1, \ldots, ht$, and then use these marginals to estimate the average of the fast states $\overline{\bz}_{t}$ which is necessary in the micro-macro solver of Eq. \eqref{eqIntx}. The complete procedure is outlined in Algorithm \ref{alg:Multi FirstScheme UKF}, with further details below.

Algorithm \ref{alg:Multi FirstScheme UKF} receives as input the prior \gls{pdf} of the fast state variables, $p(\bz_0)$. In the initialization step, the prior \gls{pdf} $p(\bz_0 | \widehat{\bz}_0, \bC_0)$ is used to generate $L = 2 d_z +1$ sigma-points and weights (for the \glspl{ukf}), as
\begin{align}
            \bz_{0}^{(i,j,0)} &= \widehat{\bz}_{0}, \quad 
            &\lambda_0^{(i,j,0)} &= \frac{1}{1+d_z},\nonumber
            \\
            \bz_{0}^{(i,j,k)} &= \widehat{\bz}_{0} + \bS_k, \quad  
            &\lambda_0^{(i,j,k)} &= \frac{1-\lambda_0^{(i,j,0)}}{2 d_z}, 
            \label{initukf} \\
            \bz_{0}^{(i,j,k+d_z)} &= \widehat{\bz}_{0} - \bS_k, \quad
            &\lambda_0^{(i,j,k+d_z)} &= \frac{1-\lambda_0^{(i,j,0)}}{2 d_z},  \nonumber
\end{align}
for $k = 1,\ldots,d_z$, where $\bS_k$ is the $k$-th row or column of the matrix square root of $\frac{d_z}{1 - \lambda_0^{(i,j,0)} } \bC_{0}$.
Therefore, in the initialization of the nested methodology (including all the layers) we obtain a set of the form $\{\btheta_0^{(i)}, \{\bx_0^{(i,j)}, \{\bz_0^{(i,j,l)}\}_{l=0}^{L-1} \}_{j=1}^J\}_{i=1}^N$, where all particles are independent at time $t=n=0$, provided the priors are independent. 

At every time step $t$, Algorithm \ref{alg:Multi FirstScheme UKF} generates new sigma-points in the space of fast states, $\{ \widetilde{\bz}_{q|h(t-1)}^{(i,j,l)}\}_{l=0}^{L-1}$, for $q = h(t-1) +1, \ldots, ht$, in step \ref{stepal3predictivez}. Note that the sigma-points are conditional on the parameters $\widetilde{\btheta}_{t}^{(i)}$ and slow variables $\bx_{t-1}^{(i,j)}$. At each time $q$, we compute a predictive mean and a covariance matrix as
\begin{eqnarray}
\widehat{\bz}_{q|h(t-1)}^{(i,j)} &=& \sum_{l=0}^{L-1} \lambda_{q-1}^{(i,j,l)} \widetilde{\bz}_{q|h(t-1)}^{(i,j,l)} \quad \text{and} \label{eqpredmeanz}
\\
{\bC}_{q|h(t-1)}^{(i,j)} &=& \sum_{l=0}^{L-1} \lambda_{q-1}^{(i,j,l)} (\widetilde{\bz}_{q|h(t-1)}^{(i,j,l)} - \widehat{\bz}_{q|h(t-1)}^{(i,j)}) (\widetilde{\bz}_{q|h(t-1)}^{(i,j,l)} - \widehat{\bz}_{q|h(t-1)}^{(i,j)} )^\top + \Delta_z \bQ_z, \label{eqpredcovz}
\end{eqnarray}
where the $\lambda_{q-1}^{(i,j,l)}$'s are the weights\footnote{These weights are deterministic and can be computed a priori in different ways. See \cite{Menegaz15} for a survey of methods.} of the sigma points ${\bz}_{q-1|h(t-1)}^{(i,j,l)}$ and $\Delta_z \bQ_z$ is the covariance matrix of the noise in Eq. \eqref{eqIntz}. The mean in Eq. \eqref{eqpredmeanz} and the covariance in Eq. \eqref{eqpredcovz} yield the approximation
\begin{equation}
\mathcal{N}(\bz_q | \widehat{\bz}_{q|h(t-1)}^{(i,j)}, {\bC}_{q|h(t-1)}^{(i,j)}) \approx p(\bz_q|\bx_{0:t-1}^{(i,j)},\by_{1:t-1},\widetilde{\btheta}_{t}^{(i)})
\label{eqX01}
\end{equation}
and we compute a new weighted set of sigma-points $\{ {\bz}_{q|h(t-1)}^{(i,j,l)}, \lambda_q^{(i,j,l)} \}$ to represent the Gaussian density in Eq. \eqref{eqX01}.

In step \ref{stepal3statespacex} of Algorithm \ref{alg:Multi FirstScheme UKF} we use the sigma-points at time $q = ht$ to generate new particles for the slow states at time $t$. Specifically, we project the ${\bz}_{ht|h(t-1)}^{(i,j,l)}$'s through the state equation of the slow state variables to obtain sigma-points in the space of the slow variables, denoted ${\bx}_{t|t-1}^{(i,j,l)}$. From these sigma-points, we obtain a mean vector and a covariance matrix, respectively,
\begin{eqnarray}
\widehat{\bx}_{t|t-1}^{(i,j)} &=& \sum_{l=0}^{L-1} \lambda_{ht}^{(i,j,l)} {\bx}_{t|t-1}^{(i,j,l)} \quad \text{and} \label{eqpredmeanx}
\\
{\bP}_{t|t-1}^{(i,j)} &=& \sum_{l=0}^{L-1} \lambda_{ht}^{(i,j,l)} ({\bx}_{t|t-1}^{(i,j,l)} - \widehat{\bx}_{t|t-1}^{(i,j)}) ({\bx}_{t|t-1}^{(i,j,l)} - \widehat{\bx}_{t|t-1}^{(i,j)} )^\top + \Delta_x \bQ_x, \label{eqpredcovx}
\end{eqnarray}
where $\Delta_x \bQ_x$ is the covariance matrix of the noise in Eq. \eqref{eqIntx}. Eqs. \eqref{eqpredmeanx} and \eqref{eqpredcovx} yield a Gaussian approximation of the predictive \gls{pdf} of the slow states, namely,
$$
p(\bx_{t}|\bx_{0:t-1}^{(i,j)},\by_{0:t-1},\widetilde{\btheta}_{t}^{(i)}) \approx \mN(\bx | \widehat{\bx}_{t|t-1}^{(i,j)},{\bP}_{t|t-1}^{(i,j)}).
$$
We generate the new particle $\widetilde{\bx}_{t}^{(i,j)}$ from this Gaussian density.

\newpage
\par\noindent\rule{0.9\textwidth}{0.3pt}
\begin{algoritmo}  {\gls{ukf} approximation of $p(\bz_{h(t-1)+1:ht} | \bx_{0:t},\by_{1:t}, \btheta)$ \label{alg:Multi FirstScheme UKF}}
	
		\textbf{Inputs}
		
		\begin{itemize}
			
			\item[-] Integration steps $\Delta_x$, $\Delta_z$ and time scale ratio $h = \frac{\Delta_x}{\Delta_z} \in \Integers^+$.
            \item[-] Prior distribution $p(\bz_0)$.
			\item[-] Known parameter vector $\widetilde{\btheta}_{t}^{(i)}$ and initial slow state $\bx_{t-1}^{(i,j)}$, i.e., fixed values for indexes $i$ and $j$.
            \item[-] If $t>1$, weighted sigma-points for the fast state at time $h(t-1)$, denoted $\{ \bz_{h(t-1)|h(t-1)}^{(i,j,l)}, \lambda_{h(t-1)}^{(i,j,l)} \}_{l=0}^{L-1}$.

		\end{itemize}

        \textbf{Initialization}	
		\begin{itemize}
        \item[-] If $t=1$, compute $L=2d_z+1$ sigma-points and weights, $\{ \bz_0^{(i,j,l)}, \lambda_0^{(i,j,l)}\}_{l=0}^{L-1}$, as in Eq. \eqref{initukf}. 
        \end{itemize}
		
		\textbf{Procedure} For $t > 1$:
		\begin{enumerate}	     
			
			\item For $l=0,\ldots,L-1$ and for $q=h(t-1)+1,...,ht$: \label{stepal3predictivez}
			
			\begin{enumerate}
			
				\item Integrate with step $\Delta_z$
				\begin{equation}
				\widetilde{\bz}_{q|h(t-1)}^{(i,j,l)} = {\bz}_{q-1|h(t-1)}^{(i,j,l)} + \Delta_z ( f_{\bz}({\bz}_{q-1|h(t-1)}^{(i,j,l)},\widetilde{\btheta}_{t}^{(i)}) + g_{\bz}(\bx_{t-1}^{(i,j)}, \widetilde{\btheta}_{t}^{(i)}) ),
				\end{equation} 
				and compute the predictive mean and covariance matrix, $\widehat{\bz}_{q|h(t-1)}^{(i,j)}$ and ${\bC}_{q|h(t-1)}^{(i,j)}$, of Eqs. \eqref{eqpredmeanz}--\eqref{eqpredcovz}. \label{stepal3propagatez}
				
				\item Approximate the predictive \gls{pdf} of the fast states as Gaussian density,
				$p(\bz_{q} | \bx_{0:t-1}^{(i,j)}, \by_{1:t-1}, \widetilde{\btheta}_{t}^{(i)}) \approx \mathcal{N}(\bz_q | \widehat{\bz}_{q|h(t-1)}^{(i,j)},{\bC}_{q|h(t-1)}^{(i,j)} )$ and use it to generate a new 
                set of weighted sigma-points $\{ {\bz}_{q|h(t-1)}^{(i,j,l)}, \lambda_q^{(i,j,l)} \}_{l=0}^{L-1}$\label{stepal3sigmapointspred}.

			\end{enumerate}

			\item In the space of the slow state variables: \label{stepal3statespacex}
			\begin{enumerate}
				
				\item Project the sigma-points ${\bz}_{h(t-1)+1:ht|h(t-1)}^{(i,j,l)}$ to obtain sigma-points in the space of the slow states,
				\begin{equation}
				{\bx}_{t|t-1}^{(i,j,l)} = \bx_{t-1}^{(i,j)} + \Delta_x(f_{\bx}(\bx_{t-1}^{(i,j)}, \widetilde{\btheta}_{t}^{(i)}) + g_{\bx}(\overline{\bz}_{t}^{(i,j,l)}, \widetilde{\btheta}_{t}^{(i)}) ),
				\end{equation}
				for $l=0,\ldots,L-1$, where $\overline{\bz}_{t}^{(i,j,l)} = \frac{1}{h} \sum_{q = h(t-1)+1}^{ht} {{\bz}_{q|h(t-1)}^{(i,j,l)}}$. Then, compute a mean vector $\widehat{\bx}_{t|t-1}^{(i,j)}$ and a covariance matrix ${\bP}_{t|t-1}^{(i,j)}$ using Eqs. \eqref{eqpredmeanx} and \eqref{eqpredcovx}.
				\label{stepal3propagatex}
				
				\item Sample $\widetilde{\bx}^{(i,j)}_{t} \sim \mathcal{N}(\bx_{t} | \widehat{\bx}_{t|t-1}^{(i,j)},{\bP}_{t|t-1}^{(i,j)})$. \label{stepal3samplex}
				
			\end{enumerate}
			
			\item Once we collect a new observation $\by_{t}$, \label{stepal3observation}
			\begin{enumerate}
			
				\item For $l = 0,\ldots,L-1$, project the sigma-points ${\bz}_{ht|h(t-1)}^{(i,j,l)}$ and the sample $\widetilde{\bx}^{(i,j)}_{t}$ into the observation space,
				\begin{equation}
				\widetilde{\by}_{t}^{(i,j,l)} = l({\bz}_{ht|h(t-1)}^{(i,j,l)}, \widetilde{\bx}^{(i,j)}_{t}, \widetilde{\btheta}_{t}^{(i)}),
				\end{equation}
				then compute the mean vector $\widehat{\by}_{ht}^{(i,j)}$ and the covariance matrix ${\bV}_{t}^{(i,j)}$ using Eqs. \eqref{eqmeany} and \eqref{eqcovy}.
				
				\item Compute $\widetilde{w}_{t}^{(i,j,l)} = p(\by_{t} | {\bz}_{ht|h(t-1)}^{(i,j,l)}, \widetilde{\bx}_{t}^{(i,j)}, \widetilde{\btheta}_{t}^{(i)}) p(\widetilde{\bx}^{(i,j)}_{t}|{\bz}_{h(t-1)+1:ht|h(t-1)}^{(i,j,l)}, {\bx}_{t-1}^{(i,j)}, \widetilde{\btheta}_{t}^{(i)})$ and the weights for the second layer
				\begin{equation}
				\widetilde{u}_{t}^{(i,j)} = \sum_{l=0}^{L-1} \lambda_{ht}^{(i,j,l)} \widetilde{w}_{t}^{(i,j,l)}.
				\end{equation}
			\end{enumerate}
			
			\item Update the  mean and the covariance matrix of the fast state variables in Eqs. \eqref{ukf:upd_mean}--\eqref{ukf:upd_cov} to obtain the new gls{pdf} $\mathcal{N}(\bz_{ht} | \widehat{\bz}_{ht|ht}^{(i,j)},{\bC}_{ht|ht}^{(i,j)})$.
			 \label{stepal3update}

			\item From the new \gls{pdf} $\mathcal{N}(\bz_{ht} | \widehat{\bz}_{ht|ht}^{(i,j)},{\bC}_{ht|ht}^{(i,j)})$, generate $L$ new sigma-points and weights $\{ \bz_{ht|ht}^{(i,j,l)}, \lambda_{ht}^{(i,j,l)}\}_{l=0}^{L-1}$. \label{stepal3sigmapointsupd}
			
		\end{enumerate}
		\textbf{Outputs:} $\{{\bz}_{ht|ht}^{(i,j,l)}\}_{l=0}^{L-1}$, $\widetilde{\bx}_{t}^{(i,j)}$, and $\widetilde{u}_{t}^{(i,j)}$.	
\end{algoritmo}
\par\noindent\rule{0.9\textwidth}{0.3pt}
\vspace{2mm}

In step \ref{stepal3observation} of the algorithm we propagate the sigma-points ${\bz}_{ht|h(t-1)}^{(i,j,l)}$ and the particle $\widetilde{\bx}_{t}^{(i,j)}$ through the observation function $l(\cdot)$ to obtain projected sigma-points (on the observation space) $\{ \widetilde{\by}_{t}^{(i,j,l)} \}_{l=0}^{L-1}$. We use these projected sigma-points to obtain a predictive mean and covariance matrix for the observation $\by_{t}$, namely,
\begin{eqnarray}
\widehat{\by}_{t}^{(i,j)} &=& \sum_{l=0}^{L-1} \lambda_{ht}^{(i,j,l)} \widetilde{\by}_{t}^{(i,j,l)} \quad \text{and} \label{eqmeany}
\\
{\bV}_{t}^{(i,j)} &=& \sum_{l=0}^{L-1} \lambda_{ht}^{(i,j,l)} (\widetilde{\by}_{t}^{(i,j,l)} - \widehat{\by}_{t}^{(i,j)}) (\widetilde{\by}_{t}^{(i,j,l)} - \widehat{\by}_{t}^{(i,j)} )^\top + \bR, \label{eqcovy}
\end{eqnarray}
where $\bR$ is the covariance matrix of the noise in the observation equation. At this step we also compute the non-normalized importance weight $\widetilde{u}_{t}^{(i,j)}$ which is output to layer 2.

In step \ref{stepal3update}, we update the mean and the covariance matrix of the fast state variables as
\begin{eqnarray}
		\widehat{\bz}_{ht|ht}^{(i,j)} &=& \widehat{\bz}_{ht|h(t-1)}^{(i,j)} + \bK_{\bz,t}^{(i,j)} \big(\by_{t} - \widehat{\by}_{t}^{(i,j)} \big), \label{ukf:upd_mean}
  \\
		{\bC}_{ht|ht}^{(i,j)} &=& {\bC}_{ht|h(t-1)}^{(i,j)} + \bK_{\bz,t}^{(i,j)} \bV_{t}^{(i,j)} \big(\bK_{\bz,t}^{(i,j)}\big)^\top, \label{ukf:upd_cov}
\end{eqnarray}
where $\bV_{t}^{(i,j)}$ is the observation covariance matrix of Eq. \eqref{eqcovy} and $\bK_{\bz,t}^{(i,j)}$ is
the Kalman gain. This one is computed as
\begin{equation}
    \bK_{\bz,t}^{(i,j)} = {\bU}_{t}^{(i,j)} \big({\bV}_{t}^{(i,j)}\big)^{-1}, \label{eqKalmangain}
\end{equation}
where the cross-covariance matrix is
\begin{equation}
{\bU}_{t}^{(i,j)} =\sum_{l=0}^{L-1} \lambda_{ht}^{(i,j,l)} ({\bz}_{ht|h(t-1)}^{(i,j,l)} - \widehat{\bz}_{ht|h(t-1)}^{(i,j)} ) (\widetilde{\by}_{t}^{(i,j,l)} - \widehat{\by}_{t}^{(i,j)} )^\top. \label{eqcrosscov}
\end{equation}
Then, in the update step, we compute the mean, $\widehat{\bz}_{ht|ht}^{(i,j)}$, and covariance matrix, ${\bC}_{ht|ht}^{(i,j)}$, of the fast state variables to obtain the approximation 
\begin{equation}
p(\bz_{ht} | \widetilde{\bx}_{t}^{(i,j)}, \bx_{0:t-1}^{(i,j)}, \by_{1:t}, \widetilde{\btheta}_{t}^{(i)}) \approx
\mN(\bz_{ht} | \widehat{\bz}_{ht|ht}^{(i,j)}, {\bC}_{ht|ht}^{(i,j)} ).
\label{eqX02}
\end{equation}
Finally, in step \ref{stepal3sigmapointsupd} we generate new weighted sigma-points to characterize the Gaussian \gls{pdf} in Eq. \eqref{eqX02}.


\section{Implementation of SMC-EnKF-EKF}\label{ap:2}


%

In Algorithm \ref{alg:Multi SecondScheme EnKF} we employ an \gls{enkf} to obtain ensemble approximations of $p(\bx_{t}|\by_{1:t-1},\widetilde{\btheta}_{t}^{(i)})$ and $p(\bx_{t}|\by_{1:t},\widetilde{\btheta}_{t}^{(i)})$. The ensembles are denoted ${\bX}_{t|t-1}^{(i)}$ and $\bX_{t|t}^{(i)}$, respectively, and they are used to approximate the computations that involved the joint \glspl{pdf} $p(\bx_{0:t}|\by_{1:t-1},\widetilde{\btheta}_{t}^{(i)})$ and $p(\bx_{0:t}|\by_{1:t},\widetilde{\btheta}_{t}^{(i)})$ in the optimal smoother. The prior distribution $p(\bx_0)$ is used to generate the initial samples utilized to build the ensemble $\bX_0^{(i)}$, $i=1, \ldots, N$.
Note that all calculations are conditional on the $i$-th parameter particle, $\widetilde{\btheta}_{t}^{(i)}$.

The structure of Algorithm \ref{alg:Multi SecondScheme EnKF} is similar to Algorithm \ref{alg:Multi FirstScheme SMC2}. We run $J$ filters of the third layer to retrieve the new samples $\widetilde{\bx}_{t}^{(i,j)}$ and obtain a predictive observation $\widetilde{\by}_{t}^{(i,j)}$ (in step \ref{stepal5newsample}). From layer 3 we also obtain the approximate likelihood $\widetilde{u}_{t}^{(i,j)}$ (in step \ref{stepal5likelihood}) and the new mean and covariance matrix of the fast state variables, $\widehat{\bz}_{ht|ht}^{(i,j)}$ and ${\bC}_{ht|ht}^{(i,j)}(\bz)$ (in step \ref{stepal5newmean}).

The non-normalized importance weight $\widetilde{v}_{t}^{(i)}$, which is output to layer 1, is computed at step \ref{stepal5likelihoodlayer1}. At step \ref{stepal5ensembles}
we generate the predictive ensemble for the slow states, ${\bX}^{(i)}_{t|t-1}$, and the observations, $\bY_{t}^{(i)}$, as well as the predictive and observation means, $\overline{\bx}_t^{(i)}$ and $\overline{\by}_t^{(i)}$. These ensembles are then used, when the new observation $\by_{t}$ is collected, to compute an updated ensemble $\bX_{t|t}^{(i)}$ (in step \ref{stepal5update}), which yields non-weighted particle approximation of the distribution with \gls{pdf} $p(\bx_{t}|\by_{1:t},\widetilde{\btheta}_{t}^{(i)})$. The update step of the \gls{enkf} can be implemented in different ways. Here we follow the scheme in \cite{Mandel06}, which avoids the direct computation of the inverse of the covariance observation matrix ($d_y \times d_y$), being better suited for high-dimensional systems. To be more specific, we compute the Kalman gain, $\bK_{\bx,t}^{(i)}$, as 
\begin{equation}
   \bK_{\bx,t}^{(i)} = {\bU}_{t}^{(i)} \big({\bV}_{t}^{(i)} \big)^{-1}, \label{eq:kalmangain_ensemble}
\end{equation}
where ${\bU}_{t}^{(i)}$ is the cross covariance matrix and $\bV_t^{(i)}$ is the covariance matrix of the observation, that are computed as
	\begin{eqnarray}
		{\bU}_{t}^{(i)} &=& \frac{1}{J}  \widetilde{\bX}_{t}^{(i)}  (\widetilde{\bY}_{t}^{(i)})^{\top}, \\
		\big({\bV}_{t}^{(i)}\big)^{-1}&=& \bR^{-1} - \bR^{-1} \frac{1}{J} \widetilde{\bY}_{t}^{(i)} \bigg(\boldsymbol{I}_J + \big(\widetilde{\bY}_{t}^{(i)}\big)^{\top} \bR^{-1} \frac{1}{J} \widetilde{\bY}_{t}^{(i)} \bigg)^{-1} \big(\widetilde{\bY}_{t}^{(i)}\big)^{\top} \bR^{-1}, \quad \quad
	\end{eqnarray}
where $\widetilde{\bX}_{t}^{(i)}={\bX}_{t|t-1}^{(i)} - \overline{\bx}_{t}^{(i)} \mathds{1}_{d_x \times J}$ and $\widetilde{\bY}_{t}^{(i)} = {\bY}_{t}^{(i)} - \overline{\by}_{t}^{(i)} \mathds{1}_{d_y \times J}$, with $\mathds{1}_{a \times b}$ describing a $a \times b$ matrix of ones.

\vspace{4mm}
\par\noindent\rule{0.9\textwidth}{0.4pt}
\begin{algoritmo} {\gls{enkf} approximation of {$p(\bx_{t} | \by_{1:t}, \btheta)$} \label{alg:Multi SecondScheme EnKF}} 
		
		\textbf{Inputs}
		\begin{itemize}
            
			\item[-] Known parameter vector $\widetilde{\btheta}_{t}^{(i)}$, i.e., the index $i$ is fixed.
            \item[-] Prior distribution $p(\bx_0)$.
            \item[-] If $t>1$, we have available the ensemble of slow states, $\bX_{t-1|t-1}^{(i)} = [\bx_{t-1}^{i,1},\ldots,\bx_{t-1}^{(i,J)}]$; the mean vector $\widehat{\bz}_{h(t-1)|h(t-1)}^{(i,j)}$ and the covariance matrix ${\bC}^{(i,j)}_{h(t-1)|h(t-1)}$, for $j= 1,\ldots,J$.

		\end{itemize}

        \textbf{Initialization}
        \begin{itemize}
            \item[-] If $t=1$, draw $J$ i.i.d. samples $\bx_0^{(i,j)}$, from the prior distribution $p(\bx_0)$, and build the ensemble $\bX_{0}^{(i)}$, as
		      \begin{equation}
		      \bX_{0}^{(i)} = [ \bx_0^{i,1}, \ldots, \bx_{0}^{(i,J)}].
		      \end{equation}
        \end{itemize}
		
		\textbf{Procedure} For $t > 0$:
		\begin{enumerate}
			\item For $j=1,\ldots,J$, run layer 3 algorithm to:
			\begin{enumerate} 
				
				\item \label{stepal5newsample} Retrieve the new sample $\widetilde{\bx}_{t}^{(i,j)}$ and obtain a predictive observation $\widetilde{\by}_{t}^{(i,j)}$.
    
                \item Compute the likelihood estimate $\widetilde{u}_{t}^{(i,j)}$. \label{stepal5likelihood}

				\item \label{stepal5newmean} Obtain the new mean $\widehat{\bz}_{ht|ht}^{(i,j)}$ and covariance matrix ${\bC}^{(i,j)}_{ht|ht}$ at time $t$. \label{stepal5propagatestate}

			\end{enumerate}

             \item \label{stepal5likelihoodlayer1} Compute the non-normalized importance weight for layer 1
				\begin{eqnarray}
				\widetilde{v}_{t}^{(i)} &=& \frac{1}{J} \sum_{j=1}^{J} \widetilde{u}_{t}^{(i,j)}.
				\end{eqnarray}
   
			\item \label{stepal5ensembles} Construct the ensembles
			\begin{equation}
			{\bX}_{t|t-1}^{(i)} = [ \widetilde{\bx}_{t}^{(i,1)}, \ldots, \widetilde{\bx}_{t}^{(i,J)} ] \quad \text{and} \quad {\bY}_{t}^{(i)} = [ \widetilde{\by}_{t}^{(i,1)}, \ldots, \widetilde{\by}_{t}^{(i,J)} ], 
			\end{equation}
			for $j= 1, \ldots, J$, and compute the prediction and observation means 
            \begin{equation}
			\overline{\bx}_{t}^{(i)} = \frac{1}{J} \sum_{j=1}^{J}\widetilde{\bx}_{t}^{(i,j)} \quad \text{and} \quad \overline{\by}_{t}^{(i)} = \frac{1}{J} \sum_{j=1}^{J} \widetilde{\by}_{t}^{(i,j)}.
			\end{equation}  
   

			\item Update the ensemble of slow variables \label{stepal5update}
           \begin{equation}
               {\bX}_{t|t}^{(i)} = {\bX}_{t|t-1}^{(i)} + \bK_{\bx,t}^{(i)} {\big(\by_{t} \mathds{1}_{d_y \times J} + \bT_{t}^{(i)} - {\bY}_{t}^{(i)}\big)},
           \end{equation}
            where $\bK_{\bx,t}^{(i)}$ is the Kalman gain of Eq. \eqref{eq:kalmangain_ensemble}, and $\bT_{t}^{(i)} = \left[ \br_{t}^{(i,1)}, \ldots, \br_{t}^{(i,J)} \right]$ with $\br_{t}^{(i,j)} \sim \mathcal{N}(\br_{t} | \boldsymbol{0}, \bR)$ is a matrix of Gaussian perturbations. 

		\end{enumerate}
		
		\textbf{Outputs:} ${\bX}_{t|t}^{(i)}$, $\{ \widehat{\bz}_{ht|ht}^{(i,j)}, {\bC}_{ht|ht}^{(i,j)} \}_{j=1}^J $ and $\widetilde{v}_{t}^{(i)}$.
		
\end{algoritmo}
\par\noindent\rule{0.9\textwidth}{0.4pt}
\vspace{2mm}


%

\newpage
Algorithm \ref{alg:Multi SecondScheme EKF} describes how to use an \gls{ekf} to obtain Gaussian approximations $\mathcal{N}(\bz_q | \widehat{\bz}_{q|h(t-1)}, {\bC}_{q|h(t-1)}) \approx p(\bz_q | \bx_{t-1}^{(i,j)}, \by_{1:t-1}, \widetilde{\btheta}_{t}^{(i)})$, $q=h(t-1)+1, \ldots, ht$, and the updated {\gls{pdf} $\mathcal{N}(\bz_{ht} | \widehat{\bz}_{ht|ht}^{(i,j)}, {\bC}_{ht|ht}^{(i,j)}) \approx p(\bz_{ht} | \widetilde{\bx}_{t}^{(i,j)},\bx_{t-1}^{(i,j)}, \by_{1:t}, \widetilde{\btheta}_{t}^{(i)})$. }We also generate the new slow states at time $t$, denoted $\widetilde{\bx}_{t}^{(i,j)}$, and the likelihood estimates $\widetilde{u}_{t}^{(i,j)} \approx p(\by_{t}|\widetilde{\bx}_{t}^{(i,j)},\bx_{0:t-1}^{(i,j)},\by_{1:t-1},\widetilde{\btheta}_{t}^{(i)})$. Note that all computations in this layer are conditional on $\bx_{t-1}^{(i,j)}$ and $\widetilde{\btheta}_{t}^{(i)}$ (i.e., the indexes $i$ and $j$ are set to a value). In the initialization step, we use the prior distribution $p(\bz_0)$ to generate the mean and covariance matrix of $\bz$, denoted $\widehat{\bz}_0^{(i,j)} = \mathbb{E}[\bz_0]$ and $\bC_0^{(i,j)}=\text{Cov}(\bz_0)$, respectively (note that they are the same for all $i$ and $j$).

 In step \ref{stepal6predictivez}, the algorithm propagates the mean, $\widehat{\bz}_{h(t-1)|h(t-1)}^{(i,j)}$, and the covariance matrix, ${\bC}_{h(t-1)|h(t-1)}^{(i,j)}$, for $q = h(t-1)+1, \ldots, ht$, conditional on the parameters $\widetilde{\btheta}_{t}^{(i)}$ and slow variables $\bx_{t-1}^{(i,j)}$. At each time step $q$, we obtain a predictive mean, $\widehat{\bz}_{q|h(t-1)}^{(i,j)}$, and the predictive covariance matrix, ${\bC}_{q|h(t-1)}^{(i,j)}$. The average of the predictive means, $\overline{\bz}_{t}^{(i,j)}=\frac{1}{h}\sum_{q=h(t-1)+1}^{ht} \widehat{\bz}_{q|h(t-1)}^{(i,j)}$, is then used to propagate the slow state $\bx_{t-1}^{(i,j)}$ and generate the new sample $\widetilde{\bx}_{t}^{(i,j)}$ from the Gaussian approximation 
$$
\mN(\bx_{t} | {\bx}_{t|t-1}^{(i,j)},\Delta_x\bQ_x) \approx p(\bx_{t}|\bx_{t|t-1}^{(i,j)},\by_{1:t-1},\widetilde{\btheta}_{t}^{(i)})
$$ 
at step \ref{stepal6samplex}. 
Note also that the covariance of the $\bz_{q|h(t-1)}^{(i,j)}$'s is neglected for simplicity in this computation and we use just the covariance matrix $\Delta_x\bQ_x$ of the slow state in Eq. \eqref{eqIntx}. 

In step \ref{stepal6observation} we project the predictive mean $\widehat{\bz}_{ht|h(t-1)}^{(i,j)}$ and the sample $\widetilde{\bx}_{t}^{(i,j)}$ into the observation space to obtain the predictive observation $\widetilde{\by}_{t}^{(i,j)}$. When the \textit{actual} observation $\by_{t}$ is available we also estimate the likelihood $p(\by_{t}|\widetilde{\bx}_{t}^{(i,j)},\bx_{0:t-1}^{(i,j)},\by_{1:t-1},\widetilde{\btheta}_{t}^{(i)})$ as 
$$
\widetilde{u}_{t}^{(i,j)} = p(\by_{t} | \widehat{\bz}_{ht|h(t-1)}^{(i,j)}, \widetilde{\bx}_{t}^{(i,j)}, \widetilde{\btheta}_{t}^{(i)}) p(\bx_{t} | \widehat{\bz}_{h(t-1)+1:ht|h(t-1)}^{(i,j)}, {\bx}_{t-1}^{(i,j)}, \widetilde{\btheta}_{t}^{(i)}).
$$
Note that we use the predictive means $\widehat{\bz}_{h(t-1):ht|h(t-1)}^{(i,j)}$ for simplicity, instead of actually integrating \gls{wrt} the random states $\bz_{h(t-1):ht}$.
Finally, in step \ref{stepal6update}, we compute the Kalman gain, $\bK_{\bz,t}^{(i,j)}$, using the predictive covariance matrix of Eq. \eqref{eqekfpredcovz} and the Jacobian matrix $\bH_{\bz,t}^{(i,j)}$. Then, we update the mean, $\widehat{\bz}_{ht|ht}^{(i,j)}$, and covariance matrix, ${\bC}_{ht|ht}^{(i,j)}$, of the fast variables for the next time step.

 
\newpage
\par\noindent\rule{0.9\textwidth}{0.4pt}
 \begin{algoritmo}  {\gls{ekf} approximation of $p(\bz_{ht} | \bx_{t},\by_{1:t}, \btheta)$ \label{alg:Multi SecondScheme EKF}}
 	
 	\textbf{Inputs}
 	\begin{itemize}
		\item[-] Prior distribution $p(\bz_0)$.
		\item[-] Integration steps $\Delta_x$, $\Delta_z$ and time scale ratio $h = \frac{\Delta_x}{\Delta_z} \in \Integers^+$.
 		
 		\item[-] Known parameter vector $\widetilde{\btheta}_{t}^{(i)}$; and known slow state vector ${\bx}_{t-1}^{(i,j)}$ (i.e. the $j$-th column of $\bX_{t-1|t-1}^{(i)}$). Namely, the indexes $i$ and $j$ are fixed.
   
        \item[-] If $t>1$, known mean fast state $\widehat{\bz}_{h(t-1)|h(t-1)}^{(i,j)}$ and covariance matrix ${\bC}_{h(t-1)|h(t-1)}^{(i,j)}$.
 	
 	\end{itemize}

 \textbf{Initialization}
 \begin{itemize}
     \item[-] If $t=1$, given the prior density $p(\bz_0) = \mathcal{N}( \bz_0| \widehat \bz_0,  \bC_0)$, we set $\widehat{\bz}_0^{(i,j)} = \widehat{\bz}_0$ and ${\bC}_{0}^{(i,j)}=\bC_0$. 
 \end{itemize}
 	
 	\textbf{Procedure} For $t > 0$:
 	\begin{enumerate}	     
 	
 		\item \label{stepal6predictivez} 
        For $q=h(t-1)+1,...,ht$, compute
 		\begin{eqnarray}
 		\widehat{\bz}_{q|h(t-1)}^{(i,j)} &=& \widehat{\bz}_{q-1|h(t-1)}^{(i,j)} + \Delta_z ( f_{\bz}(\widehat{\bz}_{q-1|h(t-1)}^{(i,j)},\widetilde{\btheta}_{t}^{(i)}) + g_{\bz}({\bx_{t-1}^{(i,j)}}, \widetilde{\btheta}_{t}^{(i)}) ) \quad \quad, \label{eqekfpredmeanz}\\ 
 		{\bC}_{q|h(t-1)}^{(i,j)} &=& \bJ_{\bz,q}^{(i,j)} {\bC}_{q-1|h(t-1)}^{(i,j)} \big(\bJ_{\bz,q}^{(i,j)}\big)^\top + \Delta_z \bQ_z, \label{eqekfpredcovz}
 		\end{eqnarray} 
 		where $\bJ_{\bz,q}^{(i,j)}$ is the Jacobian matrix of the transition function of $\bz$ and $\Delta_z \bQ_z$ is the covariance matrix of the noise in Eq. \eqref{eqIntz}.

 		\item In the space of the slow state variables:  \label{stepal6samplex}
 		\begin{enumerate}
 			
 			\item Project the predictive means $\widehat{\bz}_{h(t-1)+1:ht|h(t-1)}^{(i,j)}$ into the space of slow variables,
 			\begin{equation}
 			\widehat{\bx}_{t|t-1}^{(i,j)} = {\bx_{t-1}^{(i,j)}} + \Delta_x(f_{\bx}({\bx_{t-1}^{(i,j)}}, \widetilde{\btheta}_{t}^{(i)}) + g_{\bx}(\overline{\bz}_{t}^{(i,j)}, \widetilde{\btheta}_{t}^{(i)}) ),
 			\end{equation}
 			where $\overline{\bz}_{t}^{(i,j)} = \frac{1}{h} \sum_{q = h(t-1)+1}^{ht} {\widehat{\bz}_{q|h(t-1)}^{(i,j)}}$. \label{stepal6statespacex}
 			
 			\item Sample $\widetilde{\bx}^{(i,j)}_{t} \sim \mathcal{N}(\bx_{t} | \widehat{\bx}_{t|t-1}^{(i,j)},\Delta_x  \bQ_x)$, where $\Delta_x  \bQ_x$ is the covariance matrix of the noise in Eq. \eqref{eqIntx}. 
 			
 		\end{enumerate}

 		\item \label{stepal6observation} When a new observation $\by_{t}$ is collected: 
	 		\begin{enumerate}
 			
 			\item Project the predictive mean $\widehat{\bz}_{ht|h(t-1)}^{(i,j)}$ and the slow state $\widetilde{\bx}_{t}^{(i,j)}$ into the observation space
    \begin{equation}
 			\widetilde{\by}_{t}^{(i,j)} = l(\widehat{\bz}_{ht|h(t-1)}^{(i,j)}, \widetilde{\bx}^{(i,j)}_{t} , \widetilde{\btheta}_{t}^{(i)}). \label{eqobsekf}
 			\end{equation}
 			
 			\item Compute $\widetilde{u}_{t}^{(i,j)} = p(\by_{t} | \widehat{\bz}_{ht|h(t-1)}^{(i,j)}, \widetilde{\bx}_{t}^{(i,j)}, \widetilde{\btheta}_{t}^{(i)}) p(\bx_{t} | \widehat{\bz}_{h(t-1)+1:ht|h(t-1)}^{(i,j)}, {\bx}_{t-1}^{(i,j)}, \widetilde{\btheta}_{t}^{(i)})$. {This is an estimate of the likelihood $p(\by_{t}|\widetilde{\bx}_{t}^{(i,j)},\bx_{0:t-1}^{(i,j)},\by_{1:t-1},\widetilde{\btheta}_{t}^{(i)})$.}
 			
 		\end{enumerate}
 		
 		\item \label{stepal6update} Update the  mean and the covariance matrix of the fast variables
 			\begin{eqnarray}
 			\bK_{\bz,t}^{(i,j)} &=& {\bC}_{ht|h(t-1)}^{(i,j)} \big(\boldsymbol{H}_{\bz,t}^{(i,j)}\big)^{\top} \bigg(\boldsymbol{H}_{\bz,t}^{(i,j)} {\bC}_{ht|h(t-1)}^{(i,j)} \big(\boldsymbol{H}_{\bz,t}^{(i,j)}\big)^{\top} + \bR \bigg)^{-1},\label{eqekfKalmangain}\\
 			\widehat{\bz}_{ht|ht}^{(i,j)} &=& \widehat{\bz}_{ht|h(t-1)}^{(i,j)} + {\bK_{\bz,t}^{(i,j)} \big( \by_{t} - \widetilde{\by}_{t}^{(i,j)}\big)} \quad \text{and} \\
 			{\bC}_{ht|ht}^{(i,j)}(\bz) &=& \bigg(\boldsymbol{I}_{d_z} - \bK_{\bz,t}^{(i,j)} \boldsymbol{H}_{\bz,t}^{(i,j)}\bigg){\bC}_{ht|h(t-1)}^{(i,j)},
 			\end{eqnarray}
 		 where $\boldsymbol{H}_{\bz,t}^{(i,j)}$ is the Jacobian matrix of function $l(\cdot,\widetilde{\bx}_{t}^{(i,j)},\widetilde{\btheta}_{t}^{(i)})$ \gls{wrt} $\widehat{\bz}_{ht|ht}^{(i,j)}$, and $\bR$ is the covariance matrix of the noise in the observation equation. {We obtain $\mathcal{N}(\bz_{ht} | \widehat{\bz}_{ht|ht}^{(i,j)},{\bC}_{ht|ht}^{(i,j)}) \approx p(\bz_{ht} | \widetilde{\bx}_{t}^{(i,j)}, \bx_{0:t-1}^{(i,j)},\by_{0:t},\widetilde{\btheta}_{t}^{(i)})$.}

 	\end{enumerate}
 	\textbf{Outputs:} $\widehat{\bz}_{ht|ht}^{(i,j)}$,${\bC}_{ht|ht}^{(i,j)}$, $\widetilde{\bx}_{t}^{(i,j)}$, and $\widetilde{u}_{t}^{(i,j)}$.
 	
\end{algoritmo}
\par\noindent\rule{0.9\textwidth}{0.4pt}
\vspace{2mm}

\bibliographystyle{abbrv}
\bibliography{bibliografia2}

\end{document}